\documentclass[preprint,review,3p,11pt,times]{elsarticle}

\usepackage{hyperref}
\usepackage[section]{placeins} 
\usepackage{amsfonts}
\usepackage{amsmath}
\usepackage{amssymb}
\usepackage{xcolor}
\usepackage{graphicx}
\usepackage{subcaption}
\usepackage{enumitem}
\usepackage{booktabs}
\biboptions{sort&compress}


\journal{arXiv}

\begin{document}

\begin{frontmatter}
\title{\Large Accelerated Integration of Stiff Reactive Systems Using Gradient-Informed Autoencoder and Neural Ordinary Differential Equation}

\author[KAIST_address]{Mert Yakup Baykan\corref{mycorrespondingauthor}}
\cortext[mycorrespondingauthor]{Corresponding author}
\ead{astromert@kaist.ac.kr}
\author[KAUST_address]{Vijayamanikandan Vijayarangan}
\author[KAIST_address]{Dong-hyuk Shin}
\author[KAUST_address]{Hong G. Im}

%
%
\address[KAIST_address]{Department of Aerospace Engineering, KAIST, Daejeon,  Republic of Korea}
\address[KAUST_address]{CERP, KAUST, Thuwal, Saudi Arabia}
%
\begin{abstract}
A combined autoencoder (AE) and neural ordinary differential equation (NODE) framework has been used as a data-driven reduced-order model for time integration of a stiff reacting system. In this study, a new loss term using a latent variable gradient is proposed, and its impact on model performance is analyzed in terms of robustness, accuracy, and computational efficiency. A data set was generated by a chemical reacting solver, Cantera, for the ignition of homogeneous hydrogen-air and ammonia/hydrogen-air mixtures in homogeneous constant pressure reactors over a range of initial temperatures and equivalence ratios. The AE-NODE network was trained with the data set using two different loss functions based on the latent variable mapping and the latent gradient. The results show that the model trained using the latent gradient loss significantly improves the predictions at conditions outside the range of the trained data. The study demonstrates the importance of incorporating time derivatives in the loss function. Upon proper design of the latent space and training method, the AE+NODE architecture is found to predict the reaction dynamics at high fidelity at substantially reduced computational cost by the reduction of the dimensionality and temporal stiffness.

\end{abstract}


\end{frontmatter}


\section{Introduction}

In the pursuit of a sustainable energy mix in the future, the combustion of carbon-free or carbon-neutral fuels will become increasingly important. Clean and efficient utilization of conventional (in the near term) and alternative (in the long term) fuels requires high fidelity computational simulation tools that allow rapid predictions of reactive systems over a wide range of operating conditions \cite{chehade2021progress, lee2010studies}. The mathematical system describing chemically reacting flows involve a large dimensionality due to a number of reactive scalar variables and a wide spectrum of time scales associated with transport and chemistry. The large dimensionality poses tremendous challenges in developing a proper computational strategy to achieve desired rapid turnaround while preserving the prediction fidelity \cite{oran1991numerical, poinsot2005theoretical}.
A common workaround to avoid the disparity between the flow and reaction time scales is to apply an operator splitting method, where transport and reaction terms are integrated separately within the integration algorithm. While the method improves the computational efficiency, splitting errors need to be properly taken care of in order to preserve the solution accuracy. 
Moreover, even after the operator splitting, the large spectrum of chemical time scales still remains in the reactive terms \cite{lanser1999analysis}. In detailed chemical kinetic mechanisms used in combustion simulations, elementary reaction rates span over a wide range of magnitudes, giving rise to a stiffness problem where the fastest and slowest chemical time scales differ by many orders of magnitude. 

A traditional method to reduce the dimensionality and temporal stiffness include steady-state and partial-equilibrium approximations \cite{williams2018combustion}, eliminating a number of reactive scalar variables based on the fast species or reactions. The framework has been automated as computer-aided reduction method (CARM) \cite{doi:10.2514/6.1999-2220}. However, this approach relies on pre-selection of eliminated species based on intuition or data, and thus its general application to large systems is limited.
Alternatively, the directed relation graph (DRG) method \cite{LU20051333} has successfully been used for developing reduced mechanisms by identifying unimportant species and reactions according to their mutual association in the chemical pathways based on a pre-set criterion. A more systematic mathematical identification of fast-slow species has been provided by the computational singular perturbation (CSP) framework \cite{lam1988basic, valorani2005higher}, where the characteristic mode separation and truncation are conducted by the eigenvalue decomposition, leading to skeletal mechanisms with reduced dimension and temporal stiffness \cite{MALPICAGALASSI2018439}. The CSP algorithm has also been applied to accelerate time integration of ordinary differential equations of reactive systems by eliminating fast exhausted modes \cite{galassi2022pycsp, MALPICAGALASSI2022110875}. 

As an alternative approach to develop reduced order models (ROM) based on either experimental or computational data, the principal component analysis (PCA) has widely been used to construct a new set of orthonormal basis vectors, referred to as the principal components, which are linear transformations of the physical state variables into a reduced dimensional latent space. The resulting latent representations serve as a system of governing equations that can be integrated at much reduced computational costs \cite{SUTHERLAND20091563}. More recently, the PCA approach has been combined with CSP for additional dynamic compression in time scales \cite{malik2024combined}.

In the era of machine learning, dimensional reduction of data for various applications are commonly done by a deep neural network (DNN) architecture, where a large input variable vector undergoes multilayer perceptron (MLP) operations into a compressed latent space. A special case of the DNN framework with nonlinear activation functions applied to a solution vector is referred to as the autoencoder (AE), which is essentially a nonlinear multilayer extension of the PCA in dimensional reduction. It has been shown that AEs can effectively retain essential information when compressing data into a latent space \cite{yu2019understanding} and have since been used in predicting the evolution of dynamical systems in the latent space \cite{agostini2020exploration}. 

In the application of DNN in describing the dynamical evolution of a physical system, the neural ordinary differential equation (NODE) method \cite{chen2018neural} has been developed. Unlike traditional discrete-step models, NODEs define the state vector as a continuous representation, which allows the adaptive time integration of the state variables. Owoyele and Pal \cite{owoyele2022chemnode}  demonstrated the ability of NODEs in capturing the unsteady ignition dynamics of homogeneous reacting systems with detailed chemistry. A recent study \cite{vijayarangan2024data} conducted a more extensive investigation of AE+NODE for homogeneous reacting systems by assessing various parameters such as the number of inner layers and latent variables as well as the training methods. The study demonstrated a significantly accelerated computational cost while preserving the same level of fidelity compared to the conventional variable-order implicit integration. Moreover, a remarkable finding was that the end-to-end training in the development of AE led to a substantial removal of temporal stiffness inherent in the original chemical kinetic models. A subsequent study \cite{vijayarangan2025understanding} provided insights into why and how such a stiffness removal is automatically achieved through the AE+NODE training process. Moreover, Kumar et al. \cite{kumara2023physics} incorporated a constraint based on elemental mass conservation to enhance accuracy.

Despite these advancements, DNN-based ROMs require large datasets to accurately capture multi-mode physics. In general, ANN training relies on an inductively biased dataset due to computational complexity, large memory requirements, or difficulties in data collection. As a result, data-driven ROMs trained this way primarily learn the most probable events present in the dataset, leading to significant errors when applied to less probable scenarios. To address this limitation, Sholokhov et al. \cite{sholokhov2023physics} proposed a latent gradient loss term into the training process, thus incorporating the physical constraints into the latent space.  The results demonstrated improved model performance, particularly in scenarios where training data is scarce. 

Motivated by these new developments, the present study aims to address the additional challenges in achieving extrapolation accuracy of data-driven reduced-order models (ROMs) involving stiff chemical systems. The approach combines latent gradients \cite{sholokhov2023physics} with the architecture introduced in our previous work \cite{vijayarangan2024data} to alleviate the uncertainties associated with the strong nonlinearity of reaction rates. To demonstrate the method, we apply the approach to a constant pressure zero-dimensional homogeneous batch reactor, using both the hydrogen-air and ammonia/hydrogen-air mechanisms (with a H$_2$ blend). Specifically, we investigate: (i) the interpolation and extrapolation accuracy of the time integration of the proposed ROM for both H$_2$-air and NH$_3$-air mechanisms, (ii) the impact of different loss functions on the evolution of specific enthalpy and ignition delay time predictions, and (iii) the effect on intrinsic timescales in the latent space.

The manuscript is organized as follows: Section \ref{sec:Methodology} outlines the computational methodology used. Section \ref{sec:Results} presents the results and discusses the key contributions of this study. Finally, Section \ref{sec:Conclusion} concludes the paper and suggests potential avenues for future research.

\section{Methodology} \label{sec:Methodology}

This section summarizes the basics of the computational methods used in the study: the autoencoders (AE), neural ODEs (NODE), the combined AE+NODE architecture, and the definitions of loss terms.

\subsection{Autoencoders}

Autoencoders (AE), usually implemented as multi-layer perceptrons (MLP), apply a non-linear transformation to map an input state vector ($X \in \mathbb{R}^{N_p}$) to a lower dimensional latent space ($\hat{X} \in \mathbb{R}^{N_L}$), where essential features of the input state are captured and where $N_p > N_L$ \cite{yu2019understanding}. The nonlinearity of the transformation, achieved by the use of nonlinear activation functions between the layers of the MLP, allows increased flexibility and efficient data compression in the projections compared to any linear projection technique. In particular, the method is more effective when the physical process is highly nonlinear, such as in the reactive systems under study.

\begin{figure}[!h]
    \centering
    \includegraphics[width=0.7\linewidth]{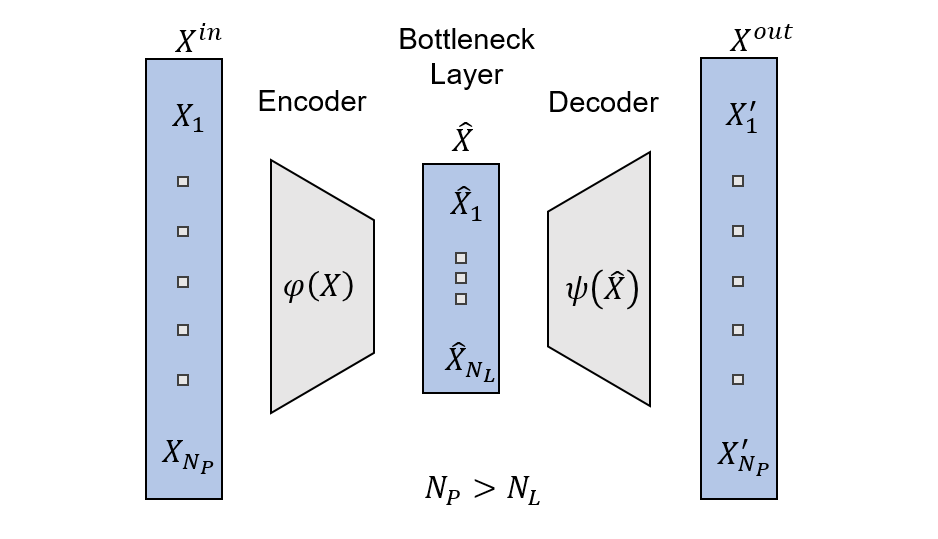}
    \caption{Schematics of an Autoencoder (AE)}
    \label{fig:autoencoder}
\end{figure}

\subsection{Neural ODEs}
Neural ODEs (NODE) are a type of deep neural network (DNN) that represents the continuous temporal dynamics of an ODE system of the form \footnote{For the present study, $X$ and $x$ are used interchangeably.}:

\begin{equation} \label{eq:node_diffeq}
 \frac{dx}{dt} = f_\theta(x, t),
\end{equation}

\noindent
where the subscript $\theta$ denotes the NODE with weights and biases that need to be optimized in order to represent the original forcing function, $f(x,t)$. The initial value problem (IVP) is then integrated from an initial condition, $x_0$, to obtain a state $x_i$ at a desired time $t_i$.

\begin{equation} \label{eq:node_diffeq2}
 x_i = x_0 + \int_{t_0}^{t_i} f_{\theta} \left( x(t), t\right) dt = \text{ODESolve}(f_\theta, x_0, (t_0, t_i)).
\end{equation}

The training to obtain the final form of $f_\theta$ is done by minimizing the scalar loss function ($L_{NODE}$) by comparing the predicted solution variable with the training data as:

\begin{equation} \label{eq:NODE_Loss}
L_\text{NODE} = \left|\left| x_\text{pred}(t) - x_\text{data}(t) \right|\right|_2^2 \,.
\end{equation}

 Unlike conventional neural network training methods, NODE uses an adjoint sensitivity method with the backpropagation algorithm. An augmented state is integrated backward in time to compute the gradients, leading to efficient training of the NODE \cite{chen2018neural}.

\subsection{AE+NODE}
\begin{figure}[!h]
    \centering
    \includegraphics[width=0.9\textwidth]{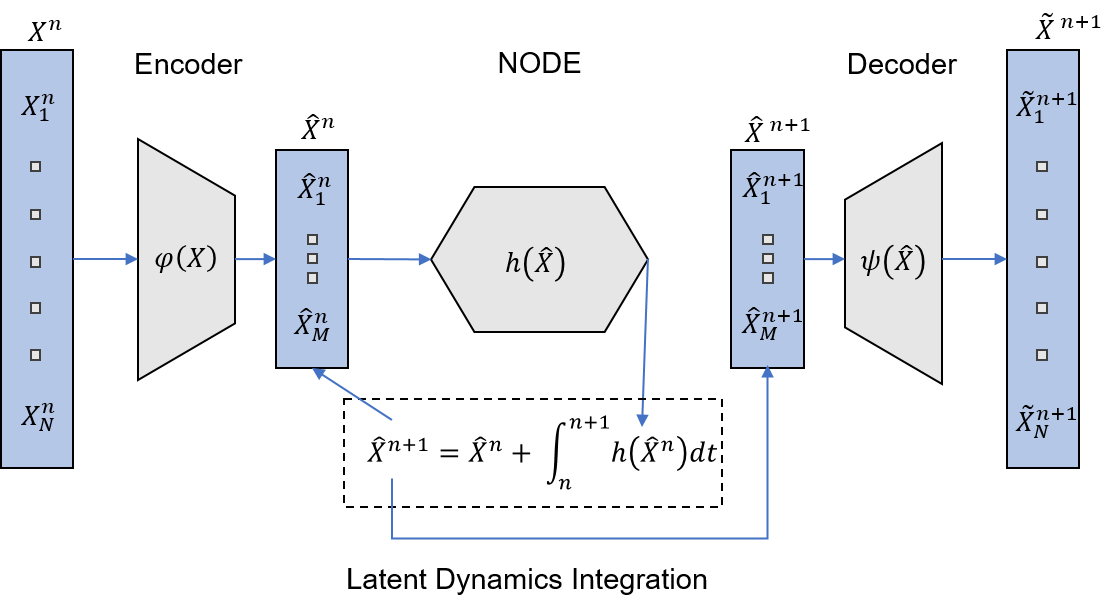}
    \caption{Schematic of AE+NODE architecture}
    \label{fig:fig1}
\end{figure}

In the present study, the combined AE+NODE architecture is used for the time integration of the ODE system. Figure \ref{fig:fig1} shows the schematic of the architecture, where the variables in the physical space ($X^n$) at the $n$-th time step are mapped to the latent space ($\hat{X}^n$) using an encoder ($\varphi(X)$).  NODE approximates the source term in the latent space $(h(\hat{X}^n))$, and  
advances the state vector from $\hat{X}^n$ to $\hat{X}^{n+1}$ using any time integration algorithm. In this work, the Runge-Kutta-Fehlberg method (RK45) was employed. Finally, the decoder ($\psi(\hat{X})$) reconstructs the variables back to the physical space ($\tilde{X}^{n+1}$) at $(n+1)$-th time step.

The physical space variables consist of temperature and mole fractions of $N_{sp}$ number of chemical species. Following the same architecture used in Ref. \cite{vijayarangan2024data}, the input vector of the dimension $N_{sp} + 1$ (=$N_p$) is compressed to the latent variable of the dimension set at 5 (= $N_L$). After the time integration, the decoder reverts the latent variable of dimension 5 back to the output vector of dimension $N_p$, although the activation functions of the encoder and decoder are not symmetric due to the nonlinearity. Both the encoder and decoder consist of 5 hidden layers with 100 neurons per layer. The bottleneck NODE consists of 5 neurons in both input and output layers and 5 hidden layers with 100 neurons each. The exponential linear activation unit is used in all the hidden layers.

The loss function ($L$) is determined by the linear combination of contributions from the encoder, neural ODE, and decoder, defined as:

\begin{equation} \label{eq:Loss}
L = \alpha_1 L_1 + \alpha_2 L_2 + \alpha_3 L_3 + \alpha_4 L_4,
\end{equation}
%
where
\begin{equation} \label{eq:L1}
L_1 = \|X - \tilde{X}\|^2_2,
\end{equation}
%
\begin{equation} \label{eq:L2}
L_2 = \|X - \psi(\varphi(X))\|^2_2,
\end{equation}
%
\begin{equation} \label{eq:L3}
L_3 = \|\hat{X} - \varphi({X})\|_1,
\end{equation}
%
\begin{equation} \label{eq:L4}
L_4 = \|h(\varphi({X})) - \nabla \varphi(X)f(X)\|^2_2,
\end{equation}
%
%
where the weights, $\alpha_i$, are constants to be adjusted to normalize the magnitude of the individual terms. 
$L_1$ is the prediction loss, which minimizes the difference between the AE+NODE prediction and the ground truth (direct solution of the original ODE). $L_2$ is the AE reconstruction loss, the discrepancy between the input and the reconstructed output produced by the autoencoder. $L_3$ is the latent variable loss between the physical and latent space. $L_2$ and $L_3$ are introduced to ensure that the encoder and decoder are bijective (one-to-one correspondence) so that the physical variables are reproduced uniquely from the latent state. $L_1$ to $L_3$ were used in the previous study \cite{vijayarangan2024data}, with $\alpha_i = 1$. 

The new addition in the present study is $L_4$, referred to as the latent gradient loss, where  $h(\varphi(X)$) is the time derivative of the latent variables obtained from the AE+NODE, and the second term is the product of the latent Jacobian, $ \nabla\varphi(X)$, and the time derivative of the physical state variables $f(X)$. The latent Jacobian is computed by
\begin{equation}\label{eq:Chemical_jacobian}
    \nabla \varphi({X})=
\begin{bmatrix}
\frac{\partial \hat{X_1}}{\partial T} & \frac{\partial \hat{X_1}}{\partial X_1} & \cdots & \frac{\partial \hat{X_1}}{\partial X_{N_{sp}}} \\
\frac{\partial \hat{X_2}}{\partial T} & \frac{\partial \hat{X_2}}{\partial X_1} & \cdots & \frac{\partial \hat{X_2}}{\partial X_{N_{sp}}} \\
\vdots & \vdots & \ddots & \vdots \\
\frac{\partial \hat{X}_{N_{l}}}{\partial T} & \frac{\partial \hat{X}_{N_{l}}}{\partial X_1} & \cdots & \frac{\partial \hat{X}_{N_{l}}}{\partial X_{N_{sp}}}
\end{bmatrix},
\end{equation}
where
\begin{equation} \label{eq:chemical_state}
X = [T, X_1, X_2, ..., X_{N_{sp}}].
\end{equation}
and the time derivative of the physical state variables ($f(X)$) are given as:
\begin{equation} \label{eq:chemical_source}
    f(X) = \frac{\partial X}{\partial t} = \left[\frac{\partial T}{\partial t}, \frac{\partial X_1}{\partial t}, \frac{\partial X_2}{\partial t}, ..., \frac{\partial X_{N_{sp}}}{\partial t}\right]^{\top}.
\end{equation}
which are evaluated by the backward difference approximation as:
\begin{equation} \label{eq:backwardDiff}
f(X^n) \approx \frac{X^{n+1}-X^{n}}{\Delta t}. 
\end{equation}

In the present study, two different loss functions are employed for the training. First, the latent variable (LV) method using
\begin{equation} \label{eq:loss_lv_definition}
L_{LV} = L_1 + L_2 + L_3,
\end{equation}
as in Ref. \cite{vijayarangan2024data}, and the latent gradient (LG) method, 
\begin{equation} \label{eq:loss_lg_definition}
L_{LG} = L_1 + L_2  + \alpha_4 L_4.
\end{equation}
where $\alpha_4 = 10^{-7}$ was selected for a proper normalization of the large magnitude of $L_4$. 
To optimize the loss functions, Adam optimizer \cite{kingma2014adam} was used with the PyTorch implementation \cite{paszke2019pytorch} in the NODE framework \cite{chen2018neural}. Losses $L_1$, $L_2$, and $L_4$ are minimized by computing the mean squared error (MSE), while $L_3$ loss is minimized by evaluating the mean absolute error (MAE), each defined as:  

\begin{equation} \label{eq:MSE}
\text{MSE} =  \frac{1}{n} \sum_{i=1}^n (X^i - \tilde{X^i})^2
\end{equation}

\begin{equation} \label{eq:MAE}
\text{MAE} =  \frac{1}{n} \sum_{i=1}^n | X^i - \tilde{X^i} |
\end{equation}

The performance of the two training methods will be compared in terms of the ignition delay time (IDT), defined as the time at the maximum $dT/dt$, and the relative root mean squared error (RRMSE) defined as:
\begin{equation} \label{eq:RRMSE}
\text{RRMSE} =  \frac{\sqrt{\frac{1}{n} \sum_{i=1}^n (X^i - \tilde{X^i})^2}}{\frac{1}{n} \sum_{i=1}^n X^i} \times 100
\end{equation}


\FloatBarrier
\section{Dataset Generation and Model Training}
\sloppy
Two fuel mixtures were considered in this study: (i) hydrogen-air mixtures and (ii) ammonia/hydrogen-air mixtures at $X_{NH_{3}}$:$X_{H_{2}} = 0.8:0.2$.  The chemical kinetic mechanisms by Marinov et al. \cite{marinov1996detailed} was used for the hydrogen-air mixtures (10 species and 27 reactions), and the mechanism of Zhang et al. \cite{zhang2021combustion}, reduced by Khamedov et al. \cite{khamedov2023structure} was used for the ammonia/hydrogen-air mixtures (25 species and 175 reactions). 

A homogeneous constant pressure reactor at 1 atm was employed to obtain the ignition curve. As the baseline ground truth, the full ODE system was integrated using Cantera \cite{cantera}. The initial temperature ($T_{0}$) ranges from 1200 K to 1400 K for the hydrogen-air mixture and 1400 K to 1600 K for the ammonia/hydrogen-air mixture, both in steps of 100 K. Equivalence ratios ($\phi$) ranging from 0.5 to 1.5 in steps of 0.02 were considered for both mixtures. The datasets from each of these conditions were saved and randomly selected for the training. The dataset was split into 85/15 for training and testing, respectively.  The dataset was further normalized by z-score normalization to accelerate the training process. Temperature and species mole fractions (except $N_2$ and Ar) were used to train the neural network. The models were trained up to $1.36\times10^{-4}$ and $2.80\times10^{-4}$ prediction accuracy ($L_1$) for the hydrogen-air and ammonia/hydrogen-air mixtures, respectively, with both LV and LG methods.

\begin{table}[!h]
\centering
\label{tab:train_conditions}
\caption{Thermodynamic conditions used to compute the constant-pressure homogeneous ignition curve data for model training and testing.}
\begin{tabular}{lcccc}
\toprule
\textbf{Mixture}       & \(\mathbf{T_\text{0} \; [\text{K}]}\) & \textbf{Equivalence ratio} (\(\boldsymbol{\phi}\)) & \textbf{Mechanism}\\ \midrule
H\(_2\)-air            & 1200 -- 1400                            & 0.5 -- 1.5            &  Marinov et al. \cite{marinov1996detailed} \\                                            & (\(\Delta T = 100\))                      & \((\Delta \phi = 0.02\)) &\\ \addlinespace 
NH\(_3\)/H\(_2\)-air   & 1600 -- 1800                             & 0.5 -- 1.5            & Zhang et al. \cite{zhang2021combustion, khamedov2023structure} \\
 ($X_{NH_{3}}$:$X_{H_{2}}$$ = 0.8:0.2 $)    & (\(\Delta T = 100\))                      & (\(\Delta \phi = 0.02\))  & \\  \bottomrule
\end{tabular}

\end{table}

\section{Results and Discussion} \label{sec:Results}
Performance of the choices of the two loss functions, the LV and LG methods, in the training process is compared in terms of the prediction accuracy within and outside the trained conditions as well as the computational efficiency. The test is conducted for the hydrogen-air and ammonia/hydrogen-air mixtures.

\subsection{Hydrogen-Air Mixture }
\FloatBarrier
\subsubsection{Accuracy in trained conditions}

\begin{figure}[hbt!]
        \centering
        \includegraphics[height=0.25\textheight]{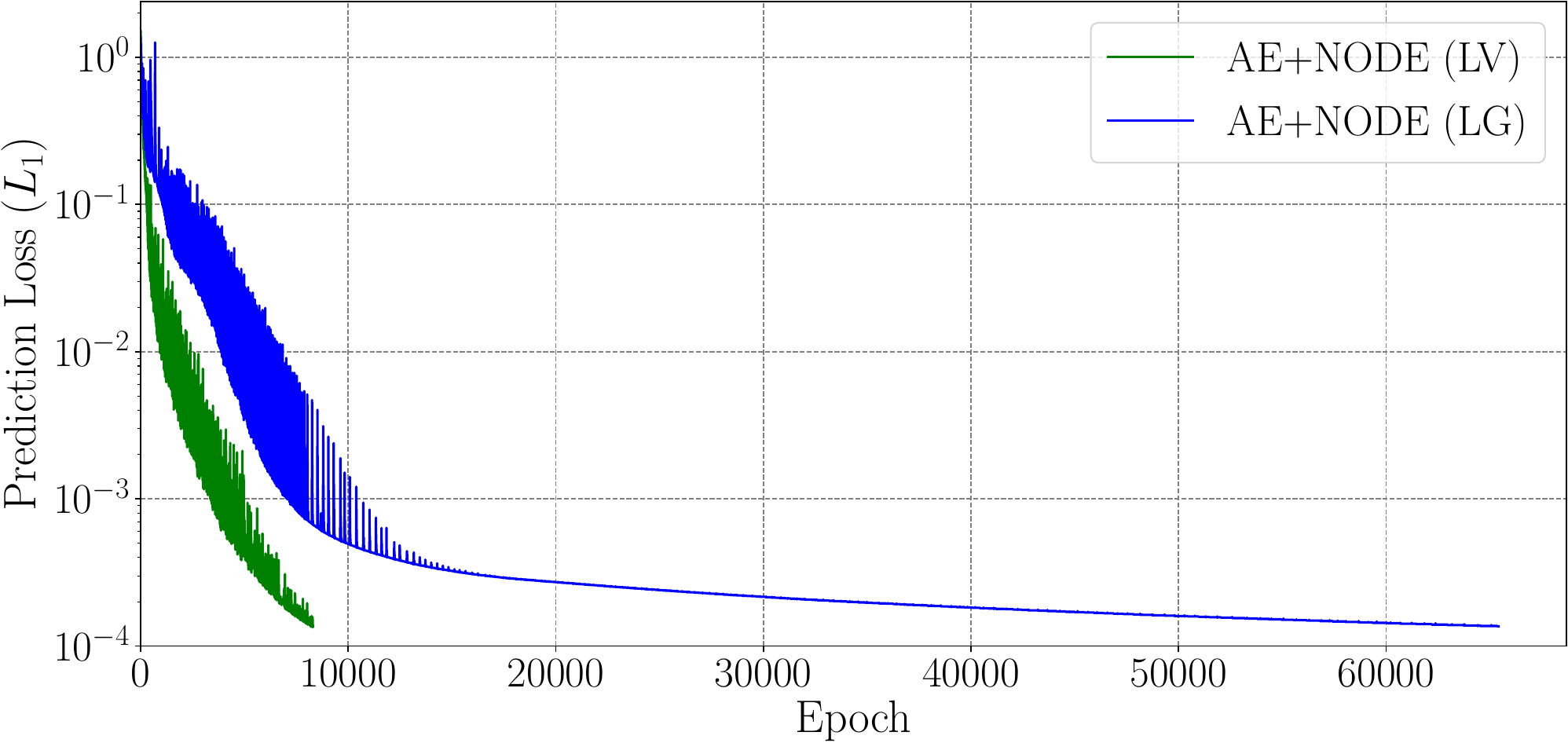}
        \caption{Prediction loss ($L_1$) curve of the models trained for hydrogen-air reaction.}
        \label{fig:H2Mech_TrainingCurve}
\end{figure}

Fig. \ref{fig:H2Mech_TrainingCurve} presents the progression of the prediction loss ($L_1$) during the training of the AE+NODE architecture using the LV and LG methods. To ensure a fair comparison, both the LV and LG methods were trained until $L_1$ loss of the test set reached $1.36\times10^{-4}$. 

The LV model achieved the desired $L_1$ test loss within 10,000 epochs, completing the process in 1.6 hours on an NVIDIA RTX3060. In contrast, the LG method required more than 65,000 epochs, resulting in a significantly longer training time of 13 hours to reach the same $L_1$ loss. Further reduction of the cut-off criterion did not lead to any substantial improvement in the results.

\begin{figure}[hbt!]
    \centering
    \begin{subfigure}[b]{0.42\textwidth}
        \centering
        \includegraphics[width=\textwidth]{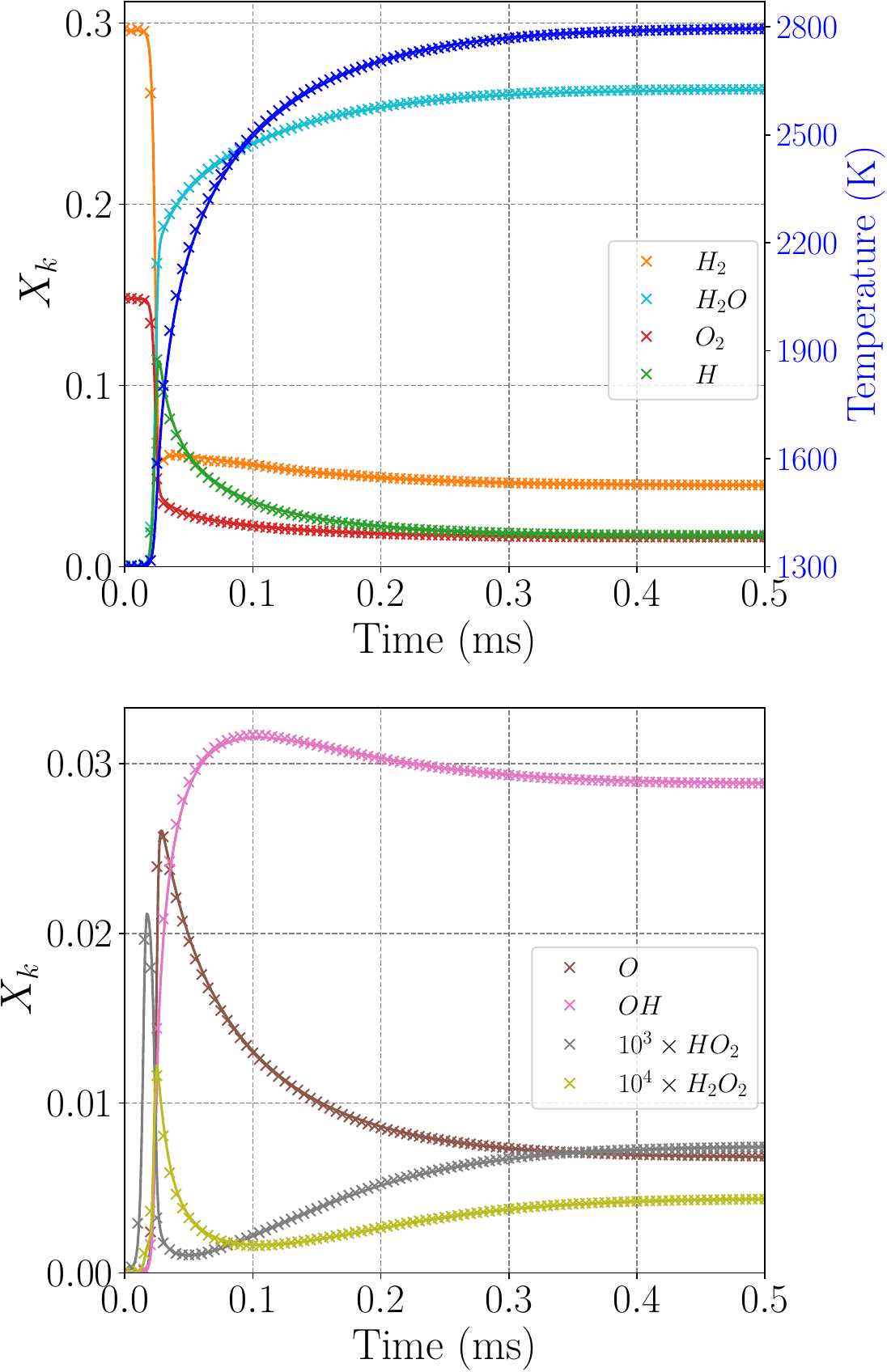}
        \caption{AE+NODE trained with LV}
        \label{fig:81019_T1300P1P1}
    \end{subfigure}
    \hspace{0.3cm}
    \begin{subfigure}[b]{0.42\textwidth}
        \centering
        \includegraphics[width=\textwidth]{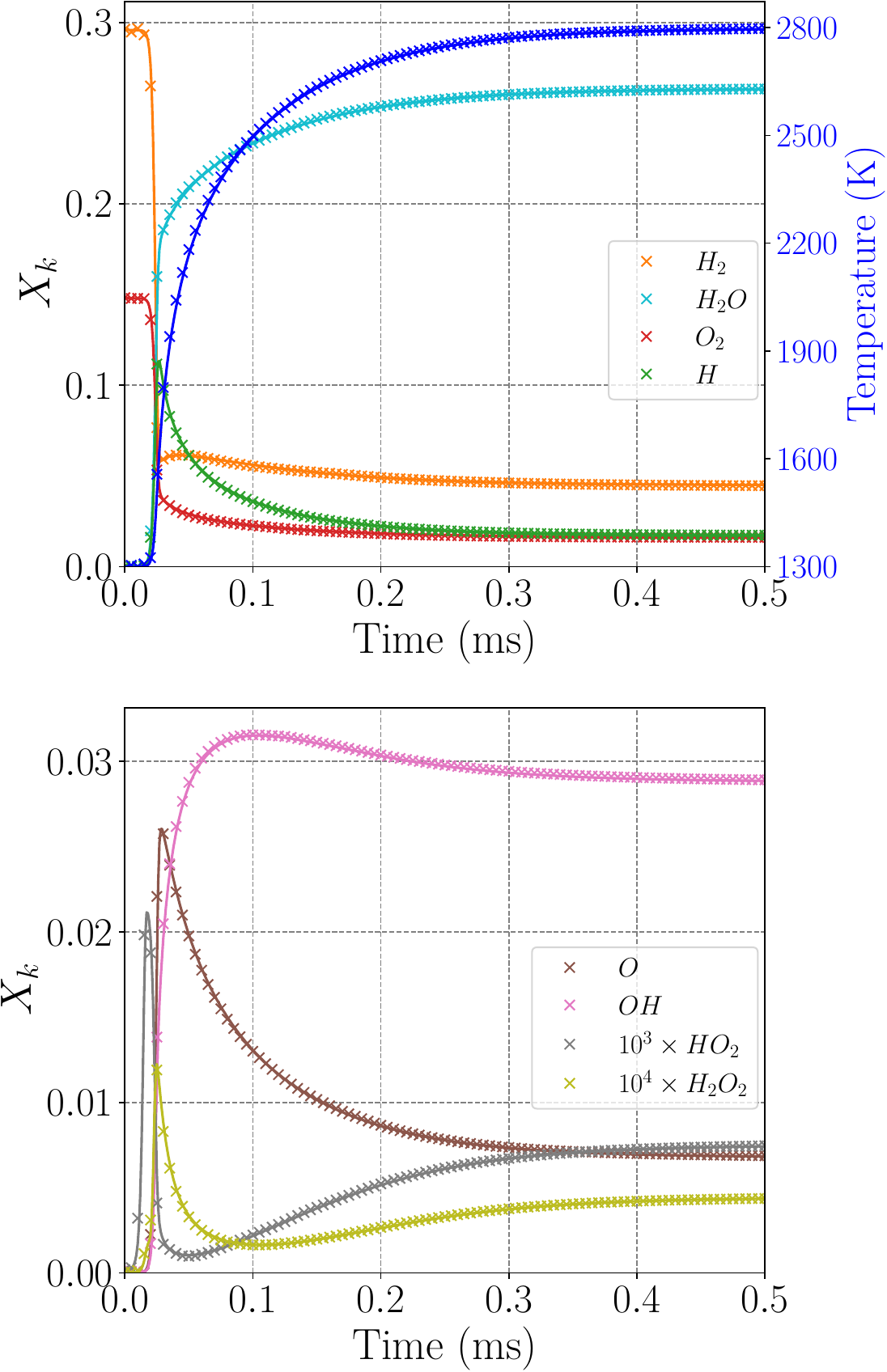}
        \caption{AE+NODE trained with LG }
        \label{fig:56908_T1300P1P1}
    \end{subfigure}
    \caption{AE+NODE (symbols) and Cantera (lines) solutions of physical state variables during constant-pressure homogeneous ignition for a trained condition (\(T_{0}\) = 1300 K, \(\phi = 1.0\)).}
    \label{fig:H2_Validation_1300K}
\end{figure}

We first examine the accuracy of models in capturing the ignition dynamics in trained conditions. Fig. \ref{fig:H2_Validation_1300K} compares the solutions of the models trained with the LV and LG methods with the solution obtained from Cantera for an initial condition $T_0 = 1300$ K and $\phi = 1.0$. As illustrated, both models accurately predict the fast-evolving ignition dynamics within the training range, precisely capturing the ignition delay time (IDT) at $t = 2.6 \times 10^{-5}$ s.

\begin{figure}[hbt!]
        \centering
        \includegraphics[height=0.25\textheight]{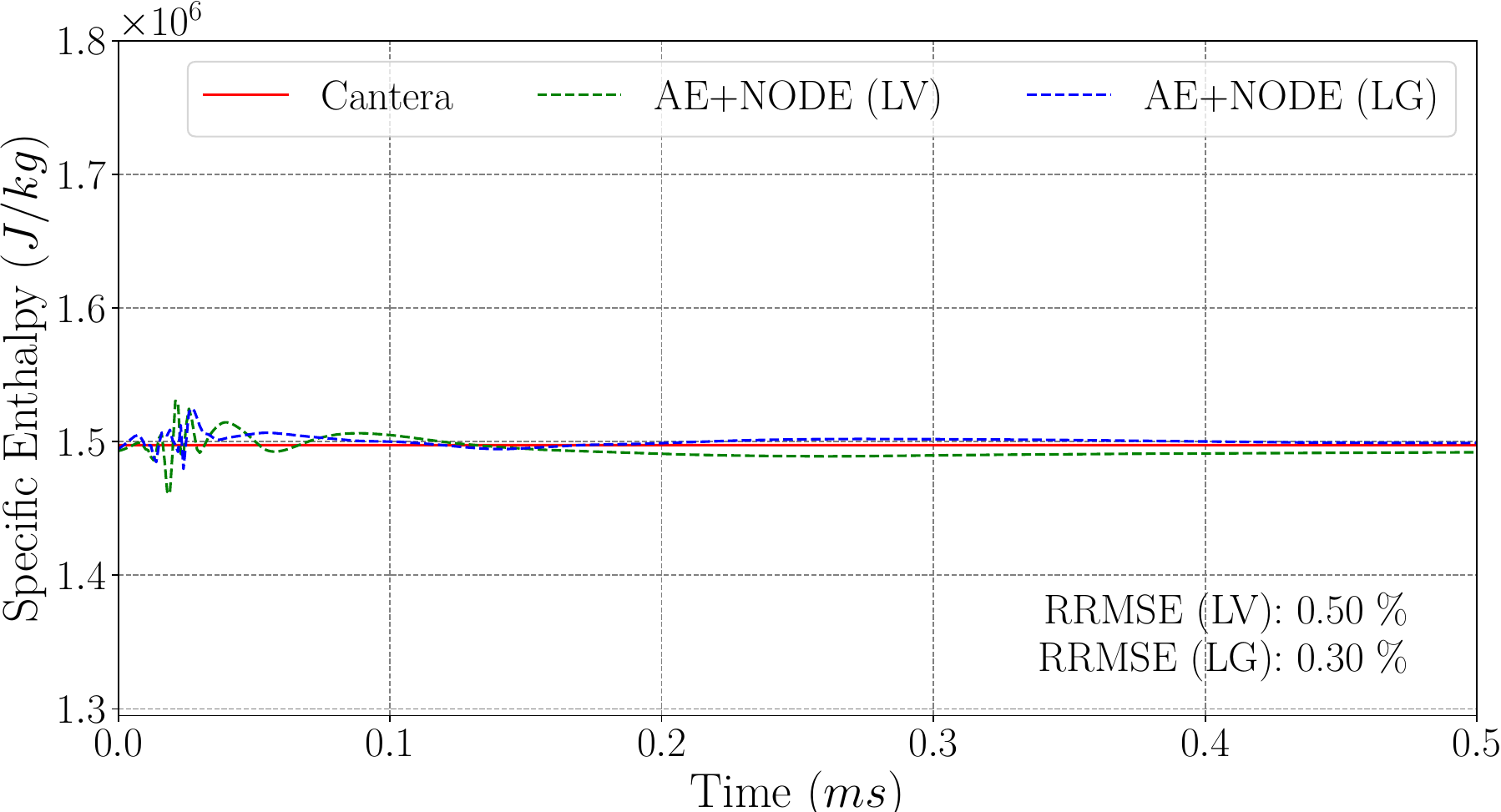}
        \caption{Evolution of the total specific enthalpy during constant-pressure homogeneous ignition ($T_{0}$ = 1300 K, $\phi$ = 1.0).}
        \label{fig:H2Mech_T1300K_Enthalpy}
\end{figure}

To further evaluate the accuracy of the AE+NODE models, the evolution of the specific enthalpy is plotted in Fig. \ref{fig:H2Mech_T1300K_Enthalpy}. For a constant-pressure adiabatic process, such as the one simulated here, specific enthalpy is expected to remain constant. We observe that both AE+NODE models effectively conserve specific enthalpy around $1.5 \times 10^{6}$ J/kg, exhibiting small errors corresponding to RRMSE values of less than 1\%. Notably, pronounced fluctuations occur around the ignition point, which can be attributed to strong chemical activity and rapid radical production occurring during this critical phase. 

\begin{figure}[hbt!]
    \centering
    \begin{subfigure}[b]{0.42\textwidth}
        \centering
        \includegraphics[width=\textwidth]{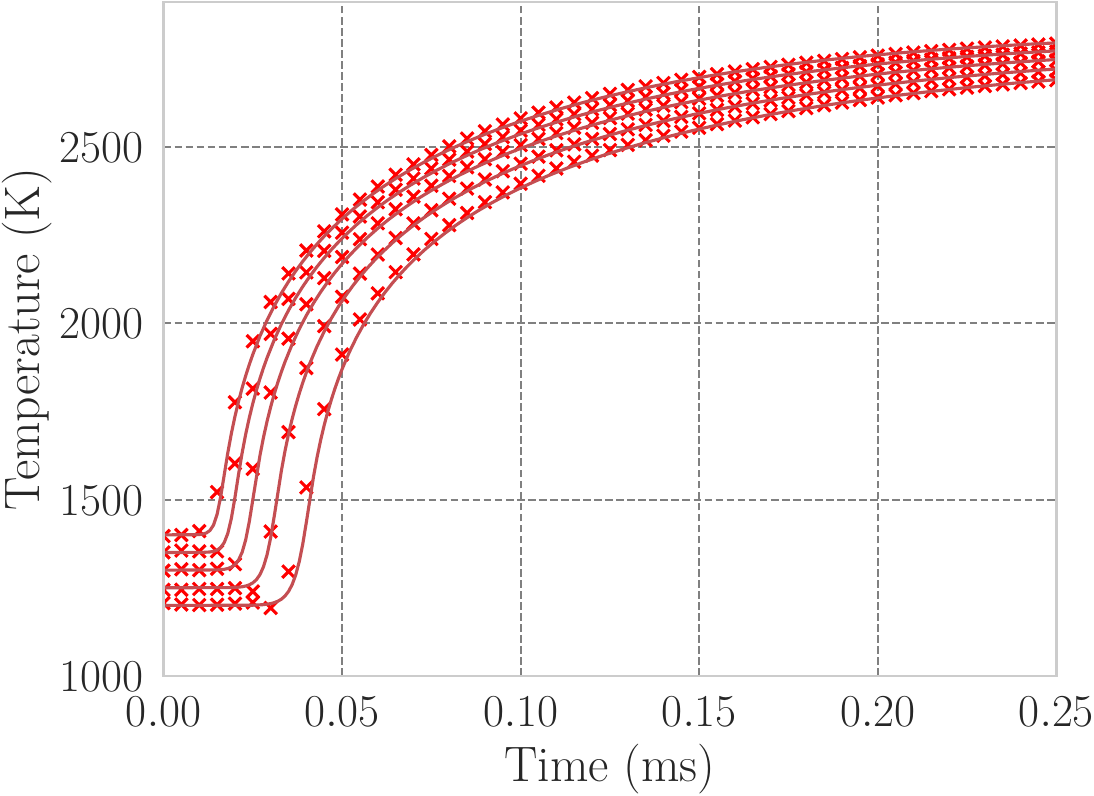}
        \caption{AE+NODE trained with LV.}
        \label{fig:81019_TempTraj}
    \end{subfigure}
    \hspace{0.3cm} 
    \begin{subfigure}[b]{0.42\textwidth}
        \centering
        \includegraphics[width=\textwidth]{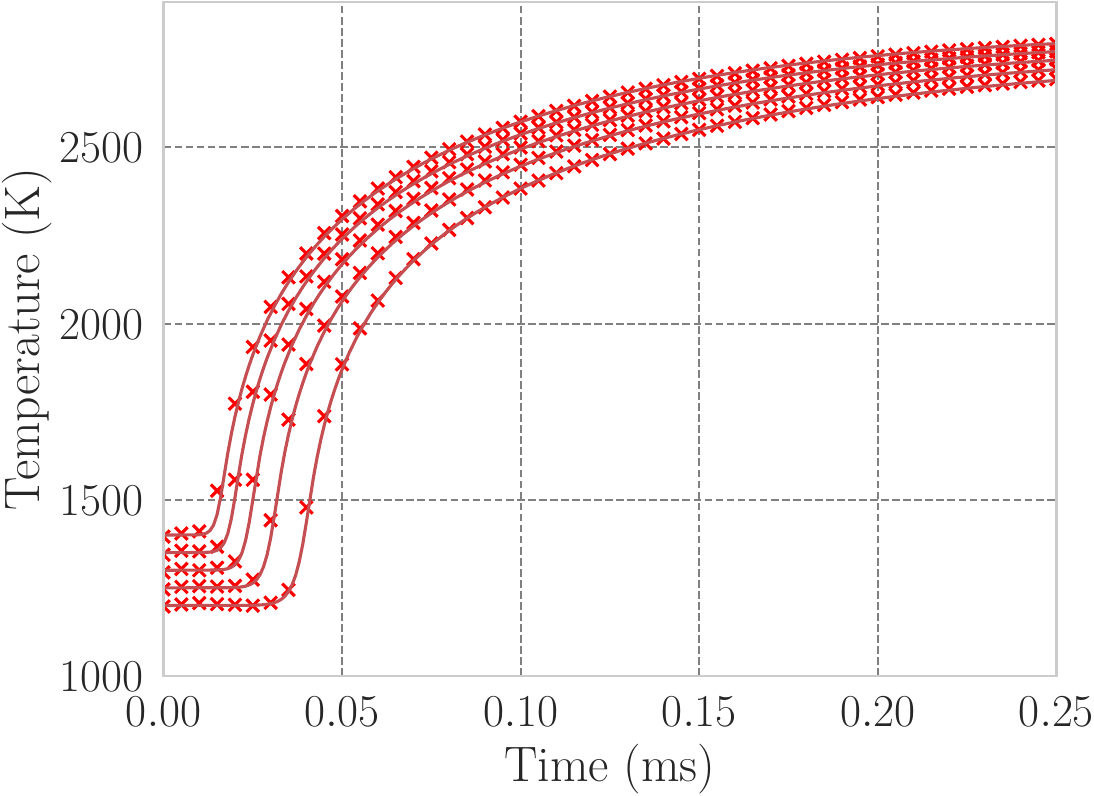}
        \caption{AE+NODE trained with LG.}
        \label{fig:56908_TempTraj}
    \end{subfigure}
        \caption{AE+NODE (symbols) and Cantera (lines) solutions of temperature evolution for various $T_0$ at $\phi = 1.0$.}
    \label{fig:H2Mech_Temp_Trajectories_Train}
\end{figure}

Predicted temperature evolutions by the LV and LG models for initial conditions within the training data range at a constant equivalence ratio are compared in Fig. \ref{fig:H2Mech_Temp_Trajectories_Train}. As anticipated, both models accurately capture the temperature evolution for the trained and interpolated conditions.

\FloatBarrier
\subsubsection{Accuracy outside trained conditions}

\begin{figure}[hbt!]
\centering
   \begin{subfigure}[b]{0.42\textwidth}
        \centering
        \includegraphics[width=\textwidth]{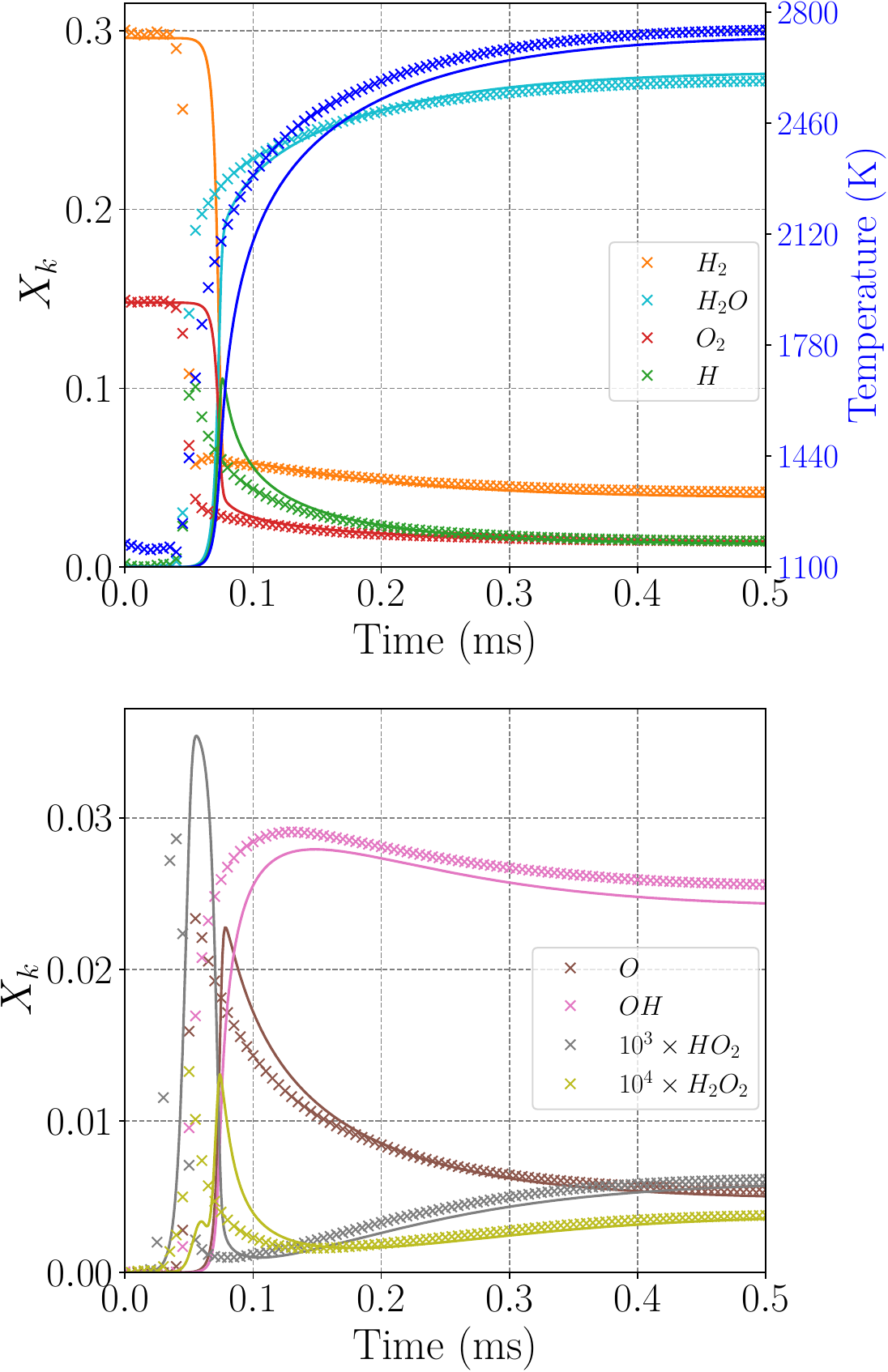}
        \caption{AE+NODE trained with LV.}
        \label{fig:H2Mech_T1100K_Comp_LV}
    \end{subfigure}
    \hspace{0.3cm} 
    \begin{subfigure}[b]{0.42\textwidth}
        \centering
        \includegraphics[width=\textwidth]{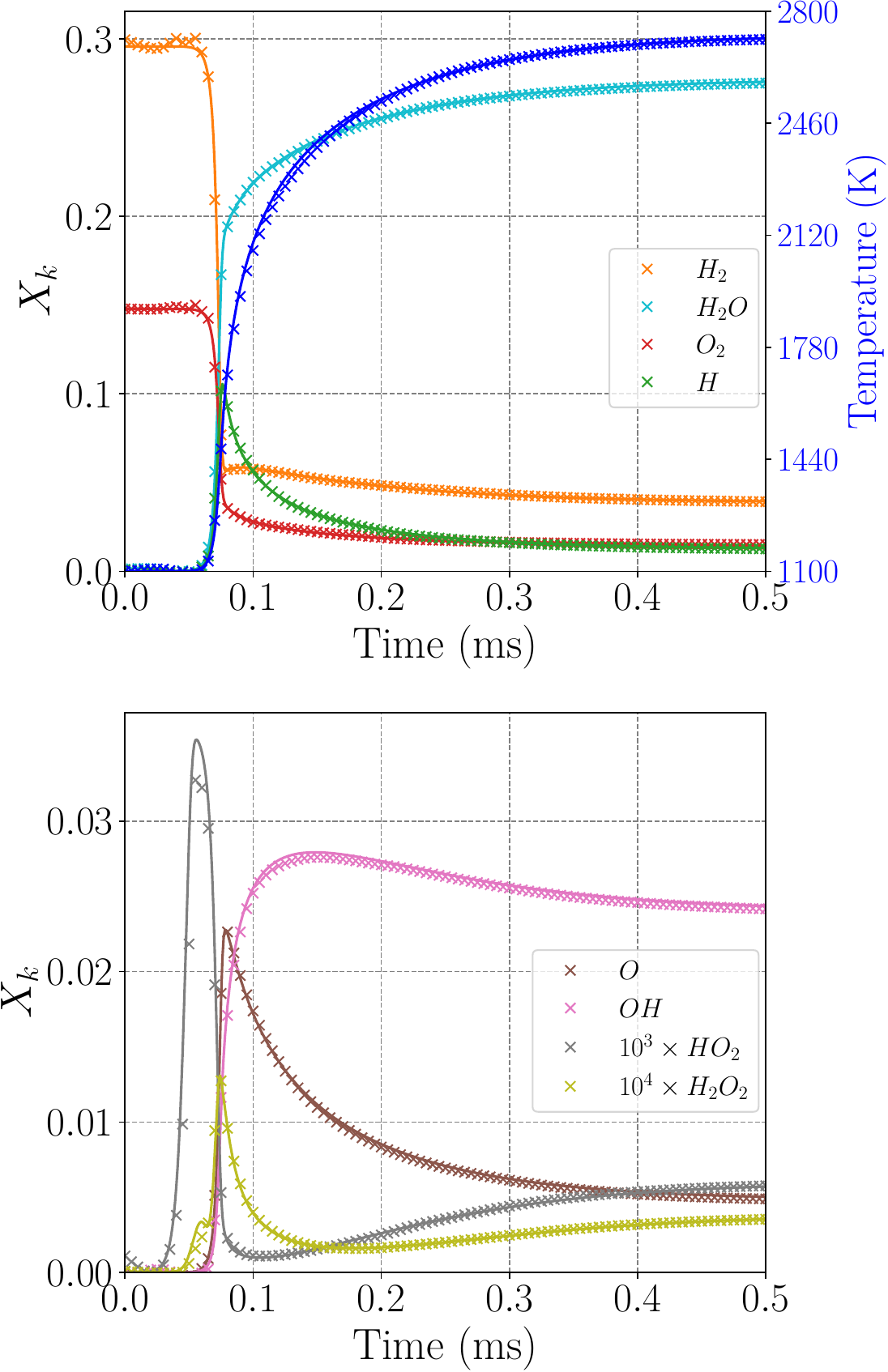}
        \caption{AE+NODE trained with LG.}
        \label{fig:H2Mech_T1100K_Comp_LG}
    \end{subfigure}
        \caption{AE+NODE (symbols) and Cantera (lines) solutions of physical state variables during constant-pressure homogeneous ignition for a condition outside the training range ($T_{0}$ = 1100 K, $\phi$ = 1.0).}
    \label{fig:H2Mech_T1100K_Comp}
\end{figure}

To assess the robustness of each model as a predictive tool, the analysis was extended to cases outside of the training range. As seen in Fig. \ref{fig:H2Mech_T1100K_Comp}, which depicts the evolution of physical state variables for an initial condition $T_0 = 1300$ K and $\phi = 1.0$, the predictions of the models vastly differ. The model trained with the LG method continues to show good agreement with the Cantera solution, accurately capturing the IDT ($t = 7.5 \times 10^{-5} \text{ s}$), the temporal evolution, and the equilibrium state of the solution variables. In contrast, the LV-trained model yields noticeable errors in the evolution of the solution variables, and in the initial temperature, which could be attributed to the cumulative errors associated with the encoding and decoding processes, as well as the poor approximation of the derivatives in the latent space. 

\begin{figure}[hbt!]
        \centering
        \includegraphics[height=0.25\textheight]{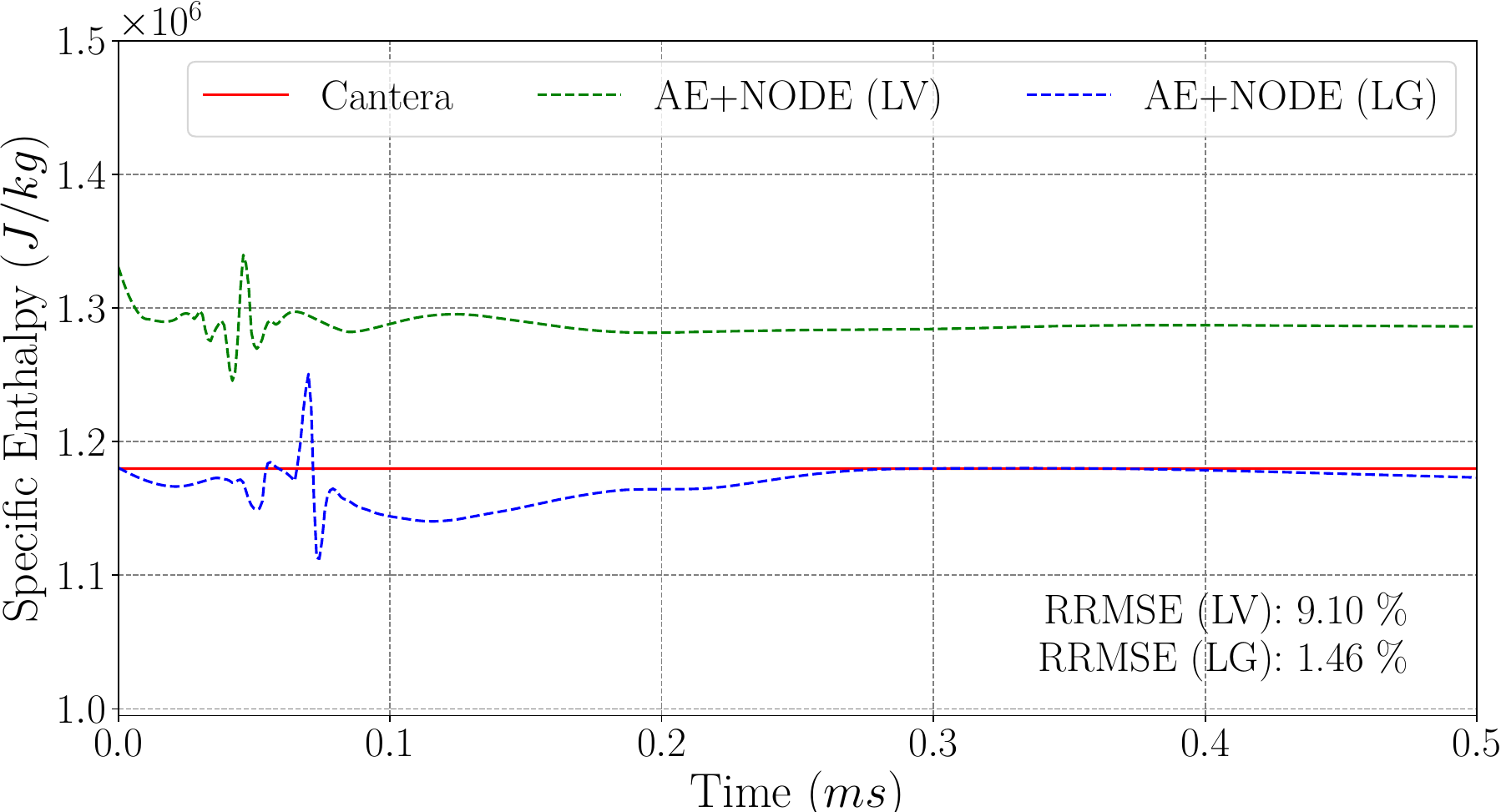}
        \caption{Evolution of the total specific enthalpy during constant-pressure homogeneous ignition (T$_{0}$ = 1100 K, $\phi$ = 1.0).}
        \label{fig:H2Mech_T1100K_Enthalpy}
\end{figure}

Examining the evolution of the specific enthalpy (Fig. \ref{fig:H2Mech_T1100K_Enthalpy}) further highlights that the LV method leads to a significantly higher error. While both models have comparable fluctuations during the ignition events, the LV-trained model has a considerably higher RRMSE (9.10\%) compared to the LG-trained model (1.46\%). The specific enthalpy exceeds the Cantera reference by approximately $1.06 \times 10^5 \text{ J}/\text{kg}$.  This error arises due to the inaccuracies in the encoding of the initial temperature. Note that the error level with the LG method is still higher than that in Fig. \ref{fig:H2Mech_T1300K_Enthalpy}, indicating the inevitable loss of accuracy in predictions for simulations with the initial conditions outside the training data range.

\begin{figure}[hbt!]
    \centering
    \begin{subfigure}[b]{0.42\textwidth}
        \centering
        \includegraphics[width=\textwidth]{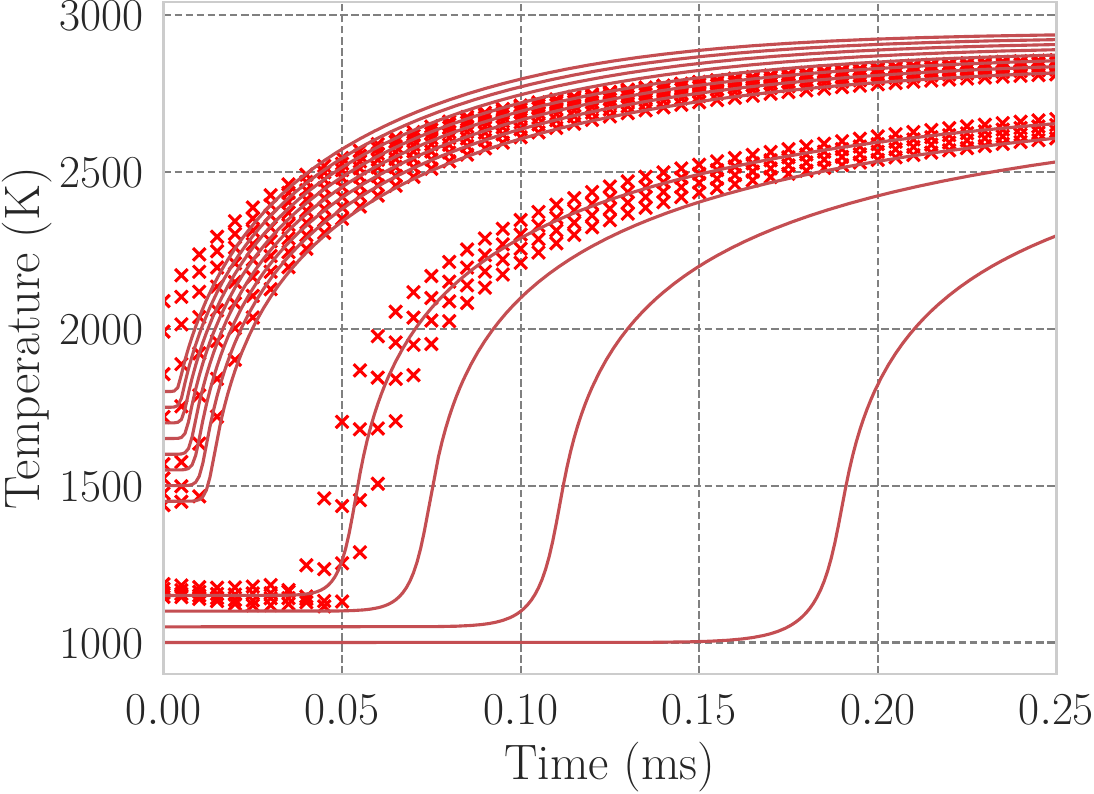}
        \caption{AE+NODE trained with LV.}
        \label{fig:81019_TempTraj_v2}
    \end{subfigure}
    \hspace{0.3cm} 
    \begin{subfigure}[b]{0.42\textwidth}
        \centering
        \includegraphics[width=\textwidth]{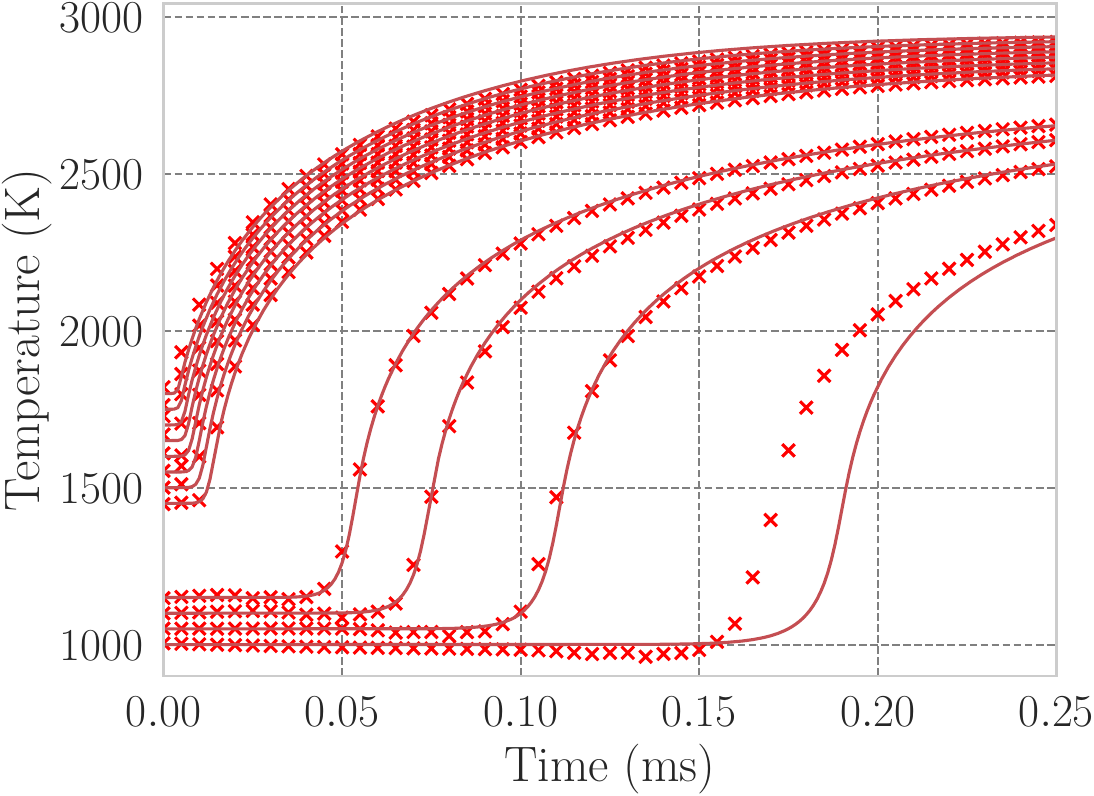}
        \caption{AE+NODE trained with LG.}
        \label{fig:56908_TempTraj_v2}
    \end{subfigure}
        \caption{AE+NODE (symbols) and Cantera (lines) solutions of temperature evolution for various $T_0$ at $\phi = 1.0$.}
    \label{fig:H2Mech_Temp_Trajectories_Test}
\end{figure}

The robustness of AE-NODE predictions outside the training data range is further examined in Fig. \ref{fig:H2Mech_Temp_Trajectories_Test} where the predicted temperature evolutions for the LV and LG models are compared at initial temperatures beyond the training range. The reference Cantera solutions are also shown in solid lines. For the LV method, predictions exhibit a consistent overestimation of the initial temperature for untrained conditions, which subsequently leads to incorrect temperature rise and the ignition delay times. In addition, the final temperature after ignition is systematically mispredicted, more concentrated over the values obtained from the data within the training range. These results highlight the LV model's limited ability to generalize to conditions far from the training data.

In contrast, the LG method demonstrates a substantially better accuracy for both initial and final temperature predictions across a wider range of conditions. The use of the latent gradient loss, $L_4$, enables the encoder/decoder to establish a more robust mapping between the initial physical and latent states, and the NODE to better approximate latent derivatives. As a result, the LG method accurately captures the initial temperature, which is critical for correctly predicting the subsequent temporal evolution of temperature, and accurately computes the latent derivatives, leading to an accurate final temperature. The errors eventually increase as $T_0$ deviates further from the training range, though to a lesser extent compared to the LV method. 

A more quantitative assessment of the prediction accuracy is shown by the RRMSE of predicted $T$ and $X_{H_2}$ by the AE+NODE models in Fig. 9. The training and untrained regions are indicated by the green and the red colors on the bottom plane, respectively. Quantitatively, RRMSE for $T$ and $X_{H_2}$ in the training range is below $1\%$ and $4\%$, respectively, for both LV and LG method. For conditions in the untrained range, however, RRMSE with the LV method reaches up to $30\%$ for $T$ and $200\%$ for $X_{H_2}$, while the LG method shows the corresponding RRMSE at much lower values of $13\%$ and $100\%$. 

The LG method clearly provides significant accuracy improvement across all unseen initial conditions, although errors do increase as conditions move further from the training data, particularly at lower $T_0$. In the untrained region, RRMSE exhibits a higher sensitivity to variations in $T_0$ compared to the equivalence ratio, which is attributed to (a) a narrower range of training data in the initial temperature range, (b) the exponential dependence of the reaction rate on temperature, and (c) the more complex chemical pathways associated with low-temperature ignition. This suggests that a more careful and extensive selection of training data with temperature as a parameter is desirable for accurate predictions.

As for the most critical quantity of interest, the errors in the IDT prediction is shown in 
Fig. \ref{fig:H2Mech_IDT_Error} for the LV and LG method, mapped over the range of initial temperature ($T_0$) and equivalence ratio ($\phi$) conditions. As expected from Fig. \ref{fig:H2Mech_Temp_Trajectories_Test}, the IDT predictions for the LV-trained model exhibit large errors, with the relative error reaching approximately 100\% as temperature deviates outside the training range, such as $T_0 < 1300 \text{ K}$ and $T_0 > 1700 \text{ K}$.

\begin{figure}[hbt!]
    \centering
    \begin{subfigure}[b]{0.48\textwidth}
        \centering
        \includegraphics[width=\textwidth]{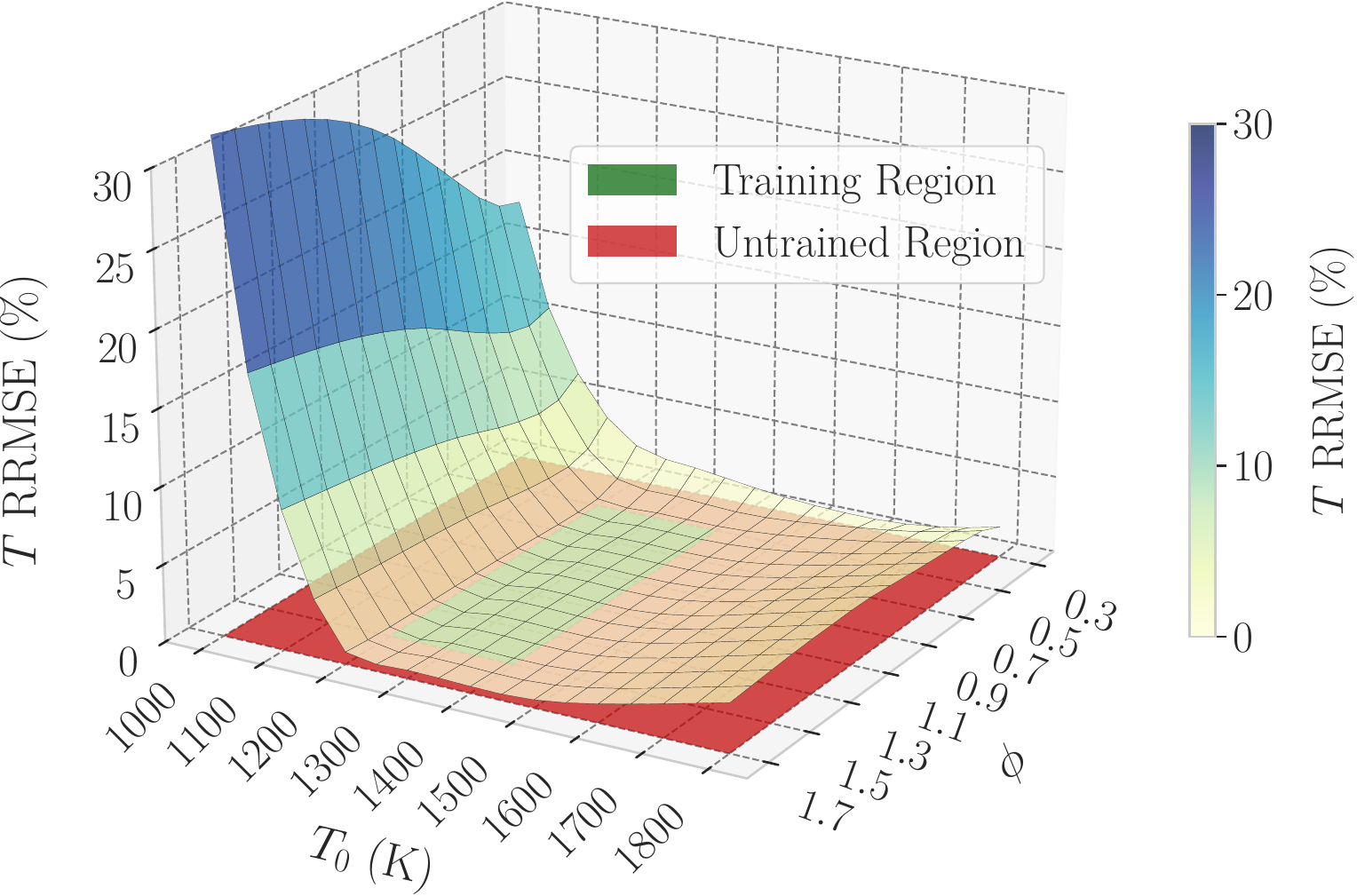}
        \caption{AE+NODE trained with LV.}
        \label{fig:80119_TErrorSurface}
    \end{subfigure}
    \hspace{0.3cm} 
    \begin{subfigure}[b]{0.48\textwidth}
        \centering
        \includegraphics[width=\textwidth]{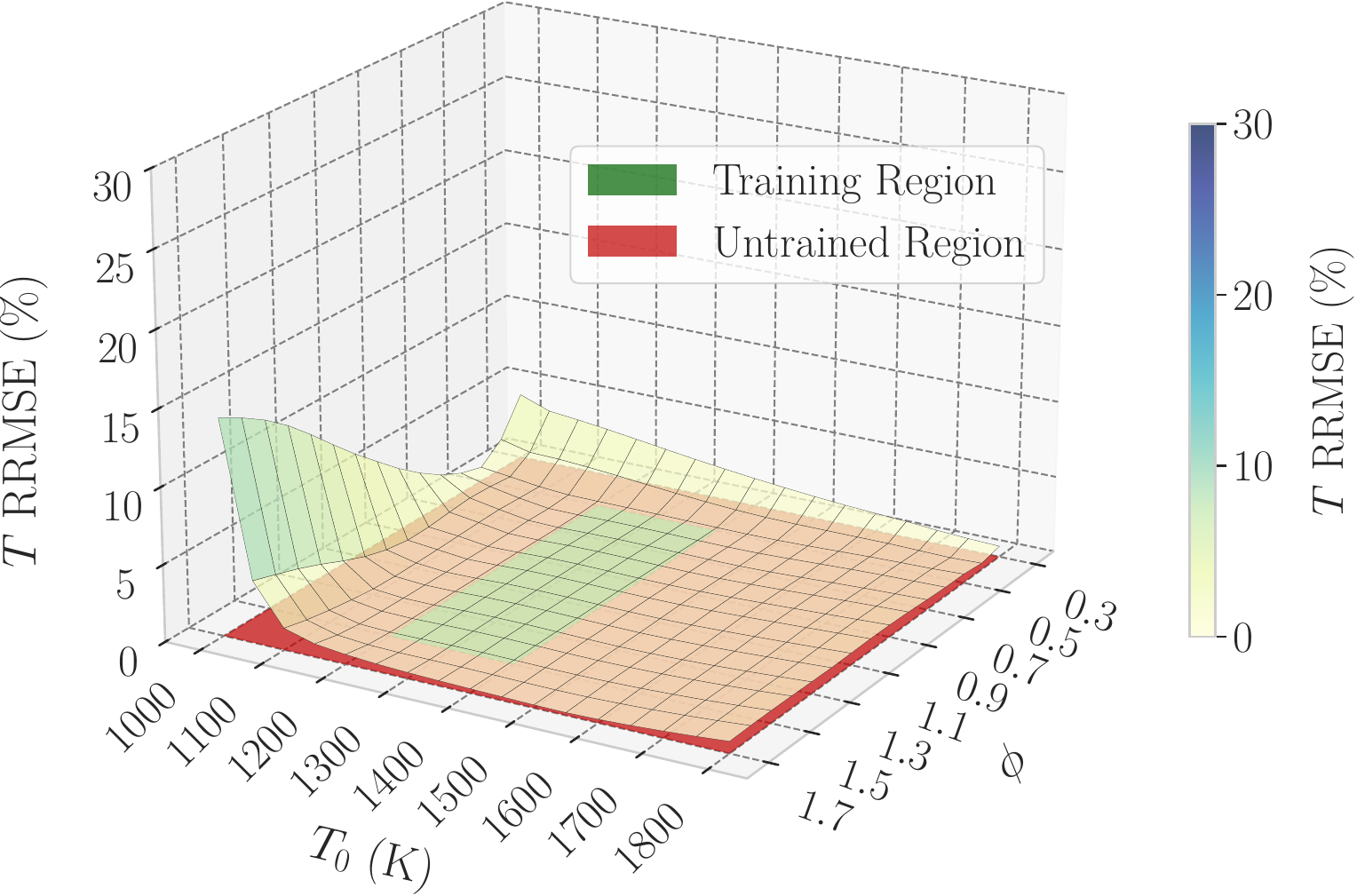}
        \caption{AE+NODE trained with LG.}
        \label{fig:56908_TErrorSurface}
    \end{subfigure}
    \par\bigskip
    \begin{subfigure}[b]{0.48\textwidth}
        \centering
        \includegraphics[width=\textwidth]{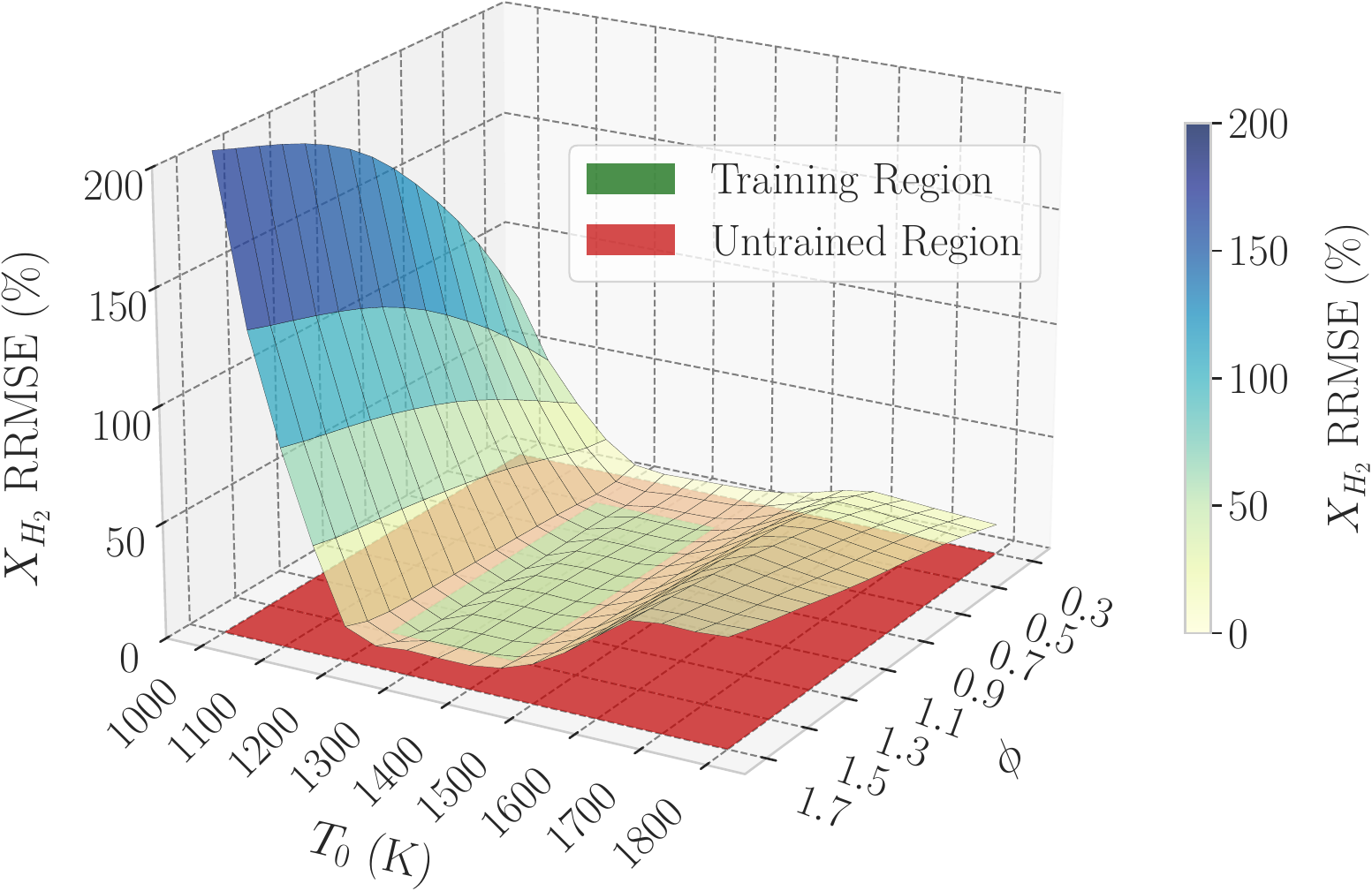}
        \caption{AE+NODE trained with LV.}
        \label{fig:81019_H2ErrorSurface}
    \end{subfigure}
    \hspace{0.3cm} 
    \begin{subfigure}[b]{0.48\textwidth}
        \centering
        \includegraphics[width=\textwidth]{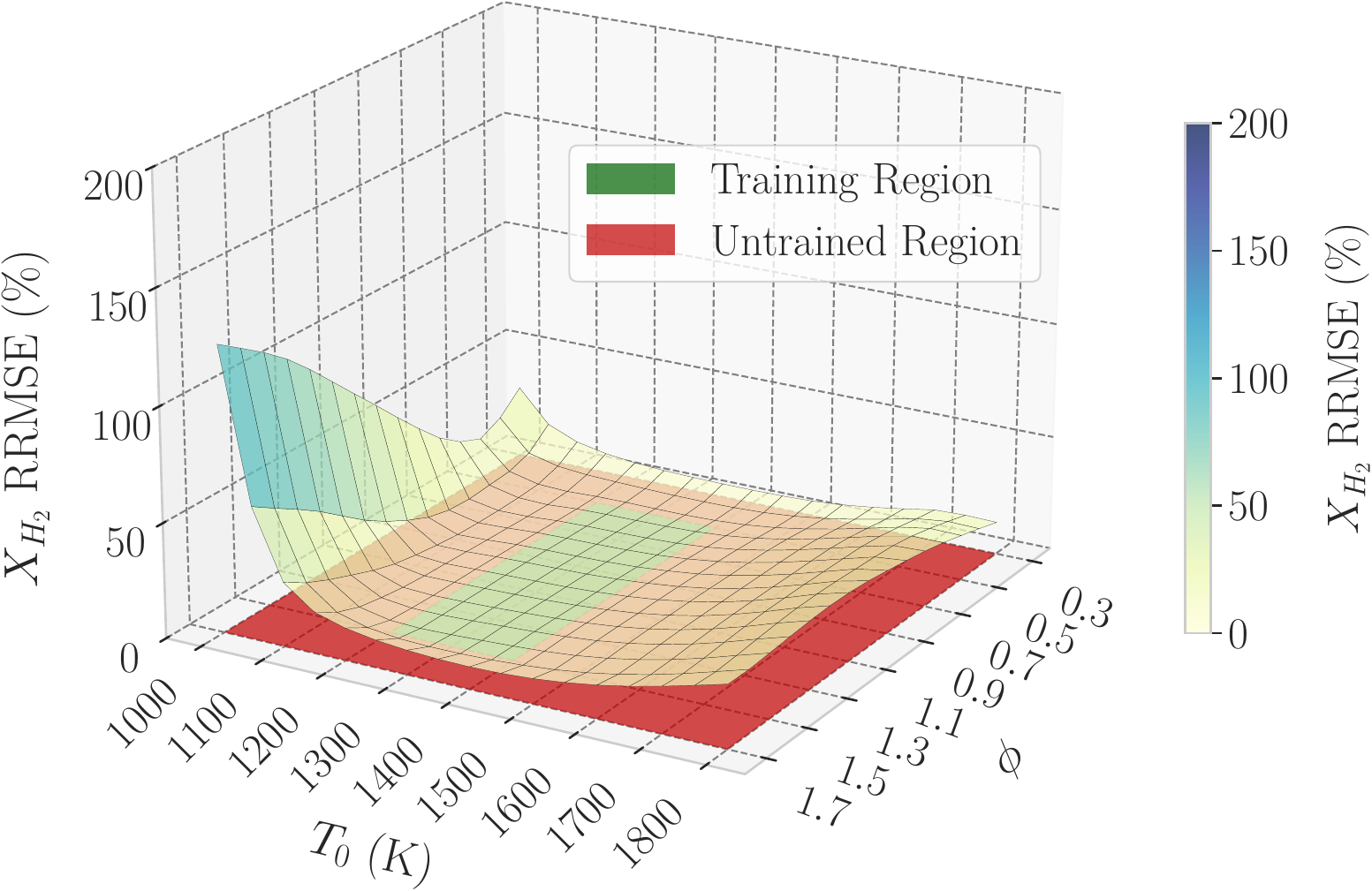}
        \caption{AE+NODE trained with LG.}
        \label{fig:56908_H2ErrorSurface}
    \end{subfigure}
        \caption{RRMSE for $T$ and $X_{H_2}$ predictions for conditions within and beyond the training range.}
    \label{fig:H2Mech_Species_RRMSE}
\end{figure}

\begin{figure}[hbt!]
    \centering
    \begin{subfigure}[b]{0.48\textwidth}
        \centering
        \includegraphics[width=\textwidth]{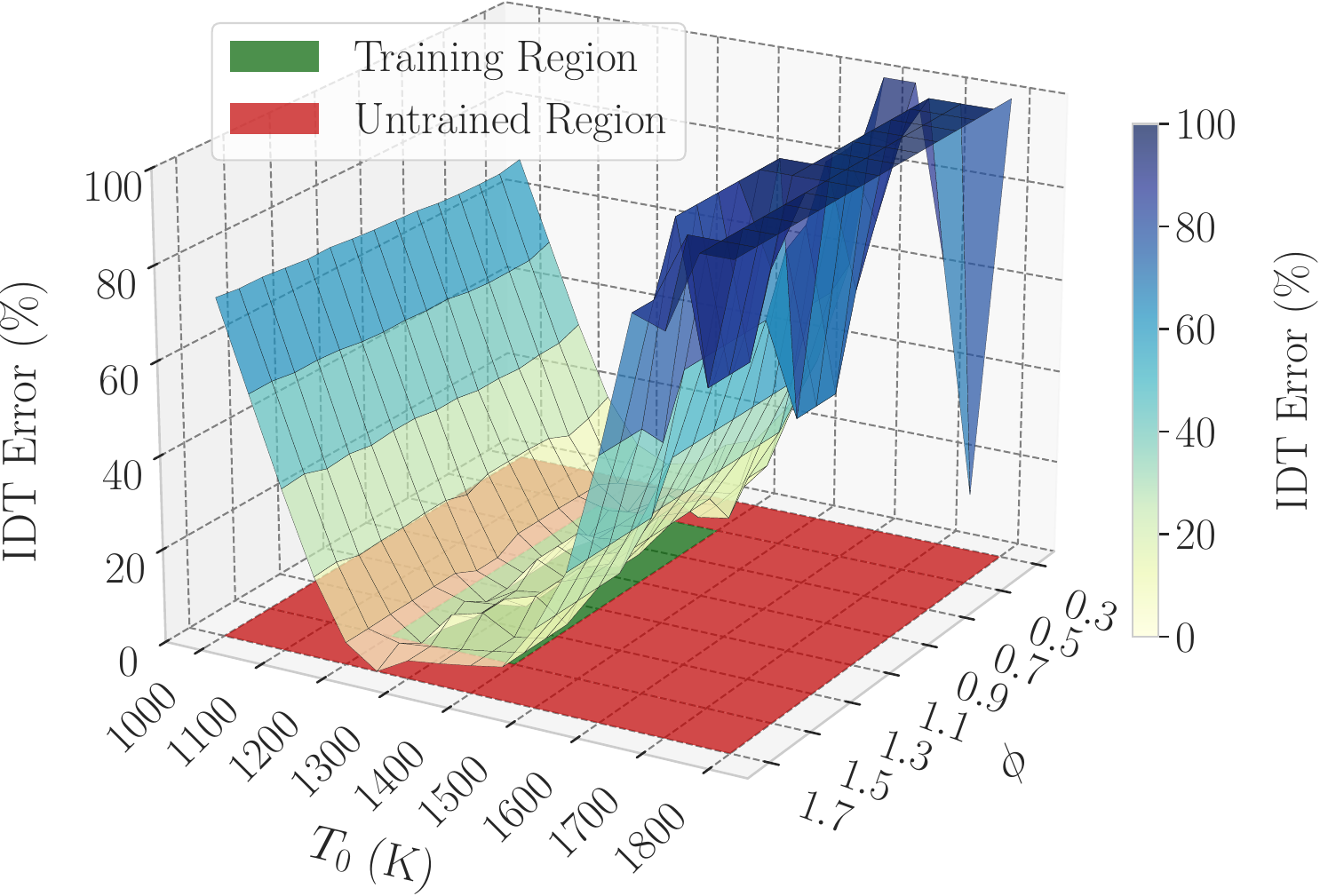}
        \caption{AE+NODE trained with LV.}
        \label{fig:81019_IDTError}
    \end{subfigure}
    \hspace{0.3cm} 
    \begin{subfigure}[b]{0.48\textwidth}
        \centering
        \includegraphics[width=\textwidth]{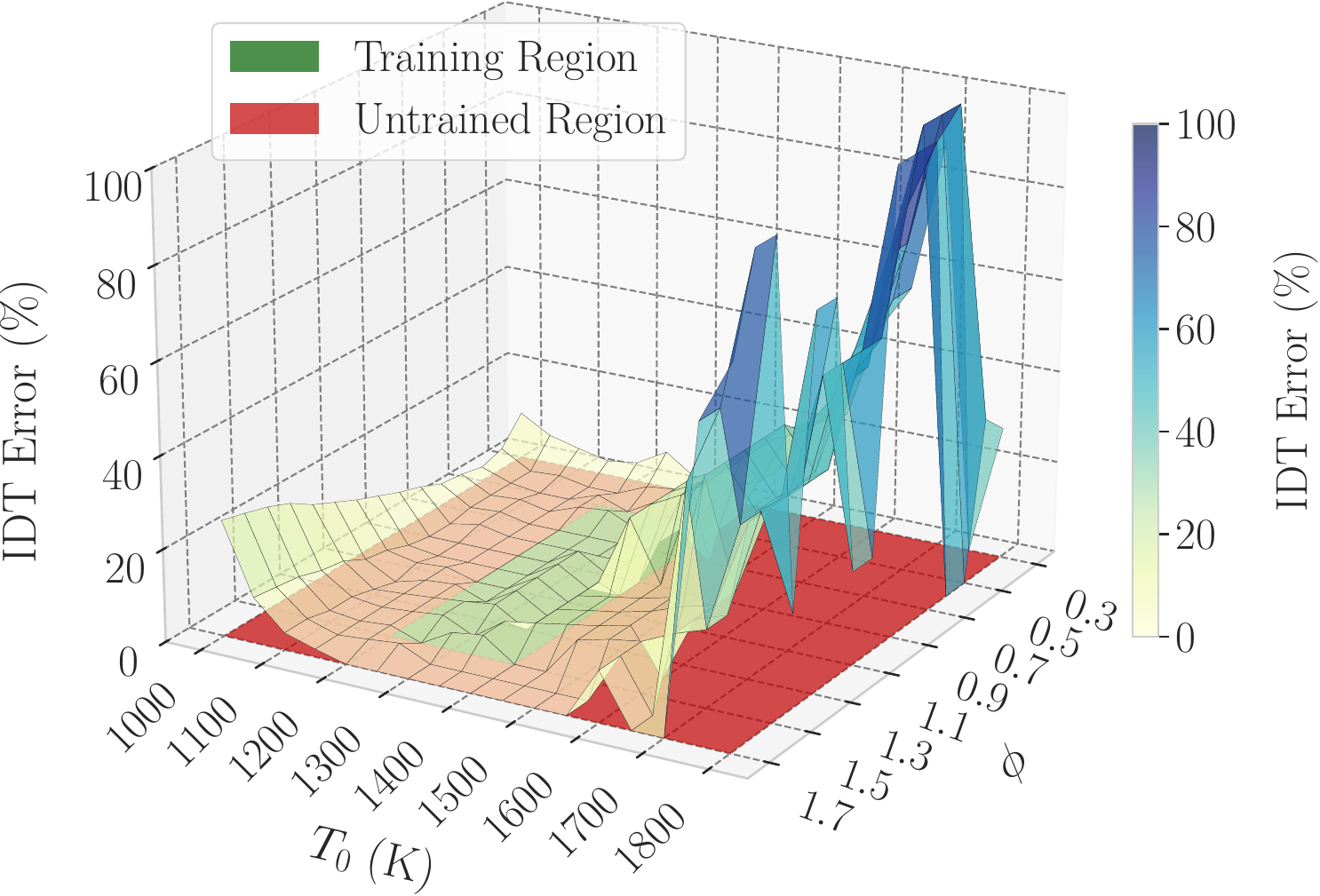}
        \caption{AE+NODE trained with LG.}
        \label{fig:56908_IDTError}
    \end{subfigure}
        \caption{IDT Error for conditions within and beyond the training range.}
    \label{fig:H2Mech_IDT_Error}
\end{figure}

In contrast, the IDT errors are significantly improved for the LG-trained model for the entire data range, while the maximum error is found at $T_0 > 1700 \text{ K}$ up to $100\%$. The increased IDT errors (in $\%$) at high initial temperatures are attributed to the fast ignition times, where even small discrepancies in the predicted ignition delay time magnify the relative error. Despite the localized increase in IDT error, the LG model maintains an RRMSE of less than 2\% for the predicted temporal evolution of temperature in the same range of conditions, as shown in \ref{fig:56908_TErrorSurface}.

\FloatBarrier
\subsection{Ammonia/Hydrogen-Air Mixture}
The accuracy assessment shown in the previous subsection is now repeated with the ammonia/hydrogen blend fuel case involving a larger number of reactive scalar variables and a complex chemical kinetic mechanism. It is of interest to confirm if the performance between the LV and LG methods is found to be consistent. 

\begin{figure}[hbt!]
        \centering
        \includegraphics[height=0.25\textheight]{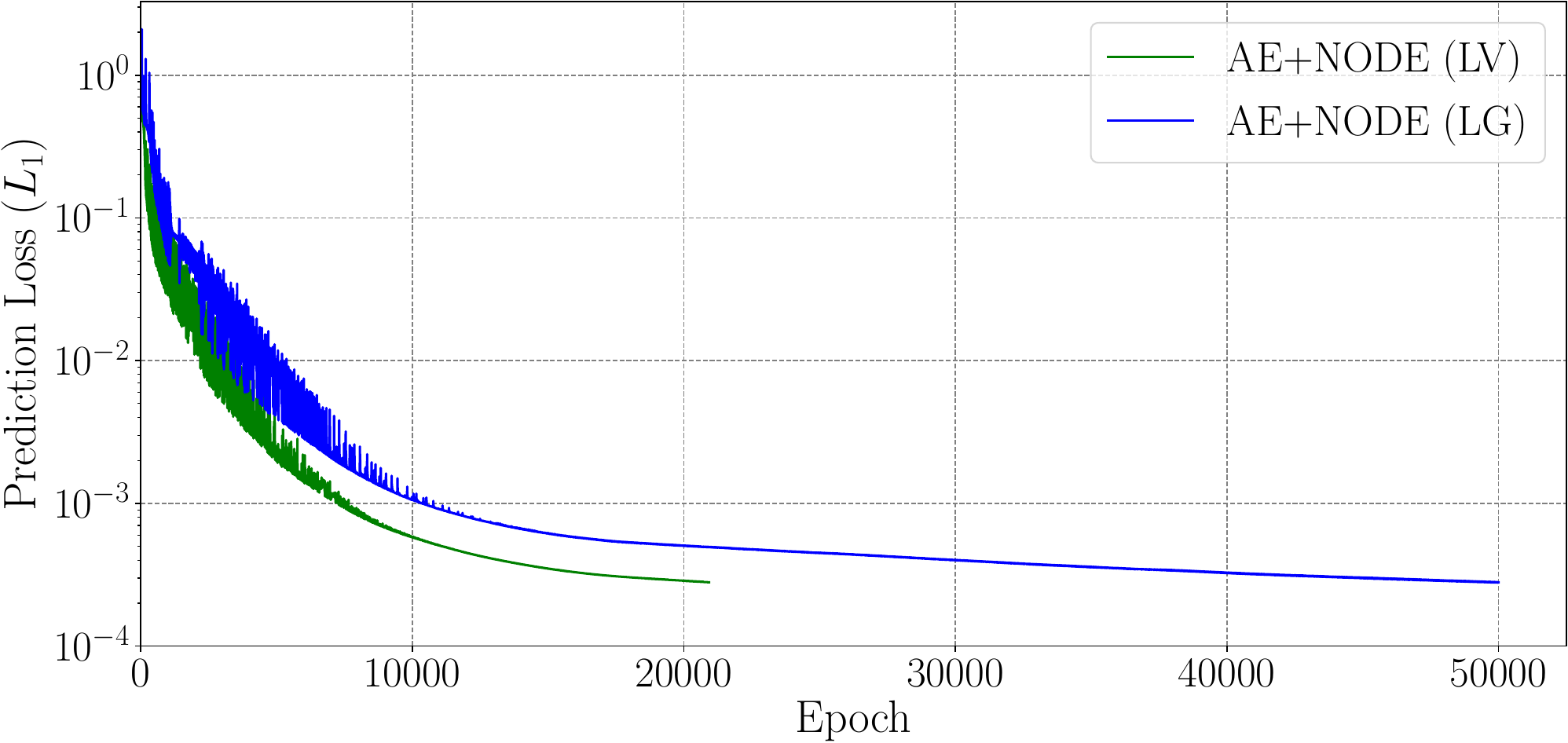}
        \caption{Prediction loss ($L_1$) curve of the models trained for ammonia/hydrogen-air reaction.}
        \label{fig:NH3Mech_TrainingCurve}
\end{figure}

Fig. \ref{fig:NH3Mech_TrainingCurve} shows the evolution of the prediction loss ($L_1$) during the training of the AE+NODE architecture with LV and LG methods. As with the hydrogen-air case, the training was stopped for both models when $L_1$ of the test set reached $2.80\times10^{-4}$, and it was also found that extending the training of the LV model toward saturation did not result in significant differences in the presented results. The training times for the models were about 6.7 hours for the LV method and 12.5 hours for the LG method on an NVIDIA RTX3060.

\FloatBarrier

The prediction accuracy with both LV and LG methods in the trained conditions is found to be accurate as in the hydrogen-air case. The results are presented in Appendix A.
For the conditions outside the trained data (Appendix B), the accuracy improvement achieved by the LG method is evident, although some  irregularities are visible in the evolution of the species mole fractions in the early stage for the ammonia/hydrogen-air case (see Fig. \ref{fig:NH3Mech_T1500K_Comp_LG}).

\begin{figure}[hbt!]
    \centering
    \begin{subfigure}[b]{0.48\textwidth}
        \centering
        \includegraphics[width=\textwidth]{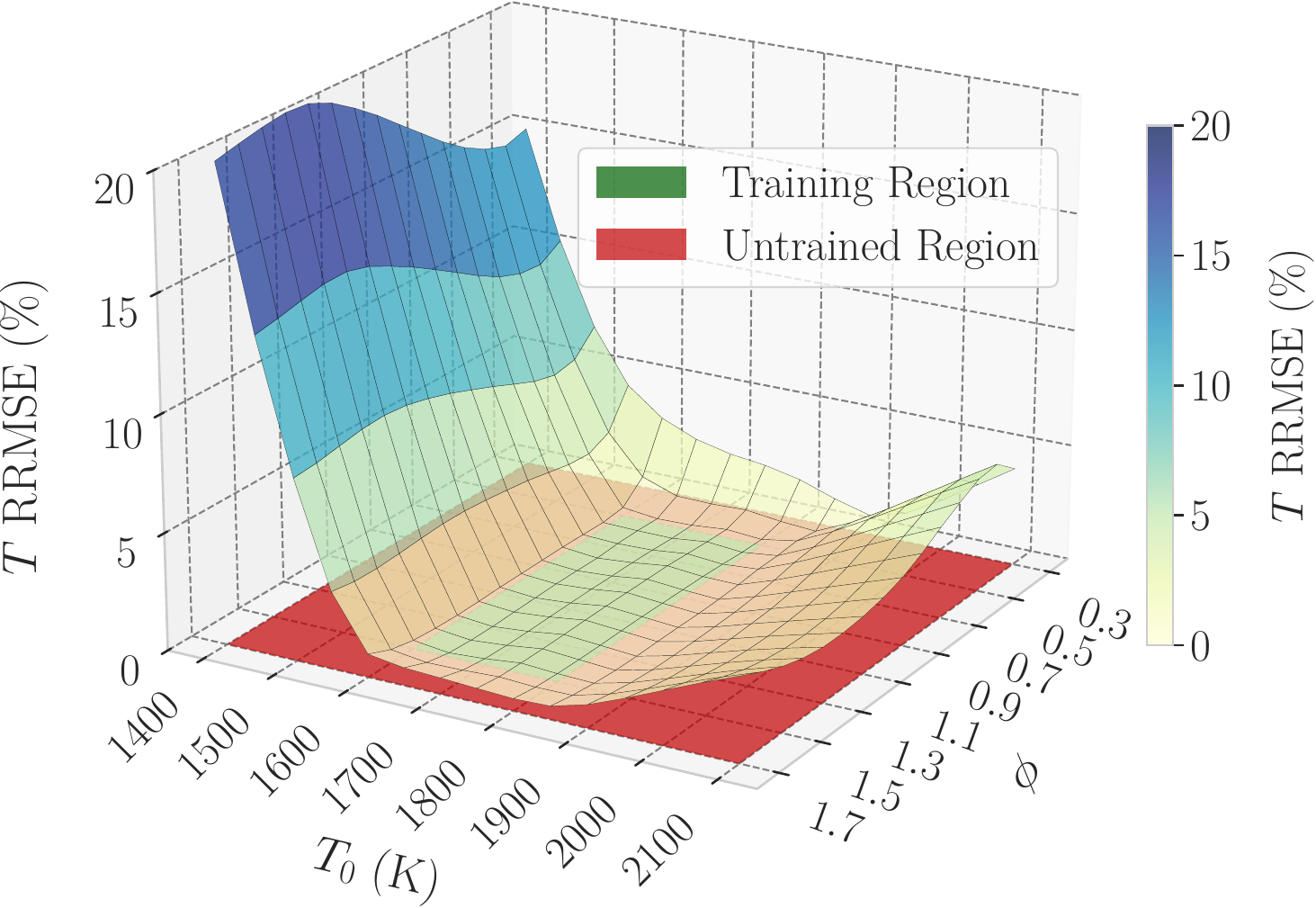}
        \caption{AE+NODE trained with LV.}
        \label{fig:48490_TErrorSurface}
    \end{subfigure}
    \hspace{0.3cm} 
    \begin{subfigure}[b]{0.48\textwidth}
        \centering
        \includegraphics[width=\textwidth]{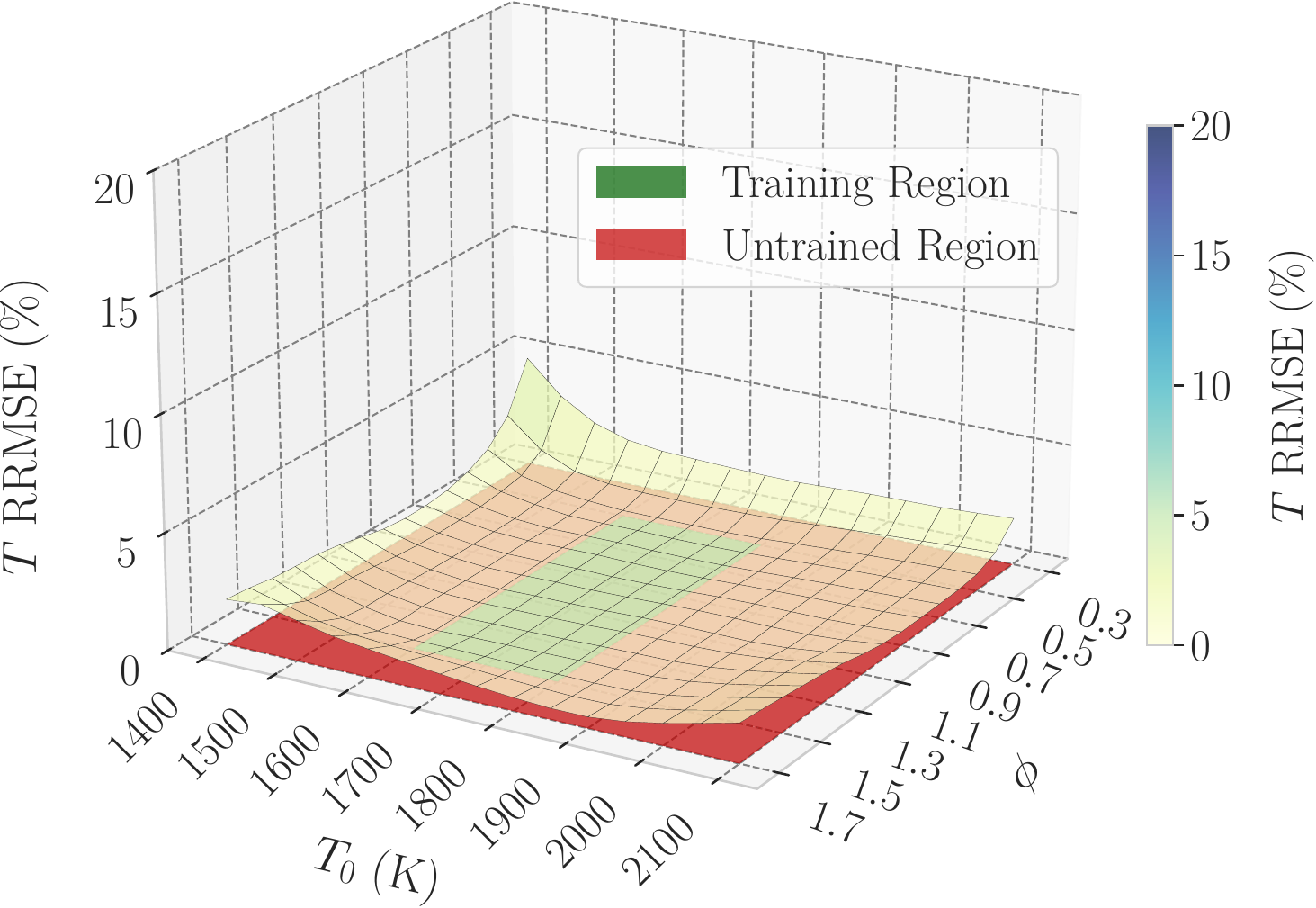}
        \caption{AE+NODE trained with LG.}
        \label{fig:37921_TErrorSurface}
    \end{subfigure}
    \begin{subfigure}[b]{0.48\textwidth}
        \centering
        \includegraphics[width=\textwidth]{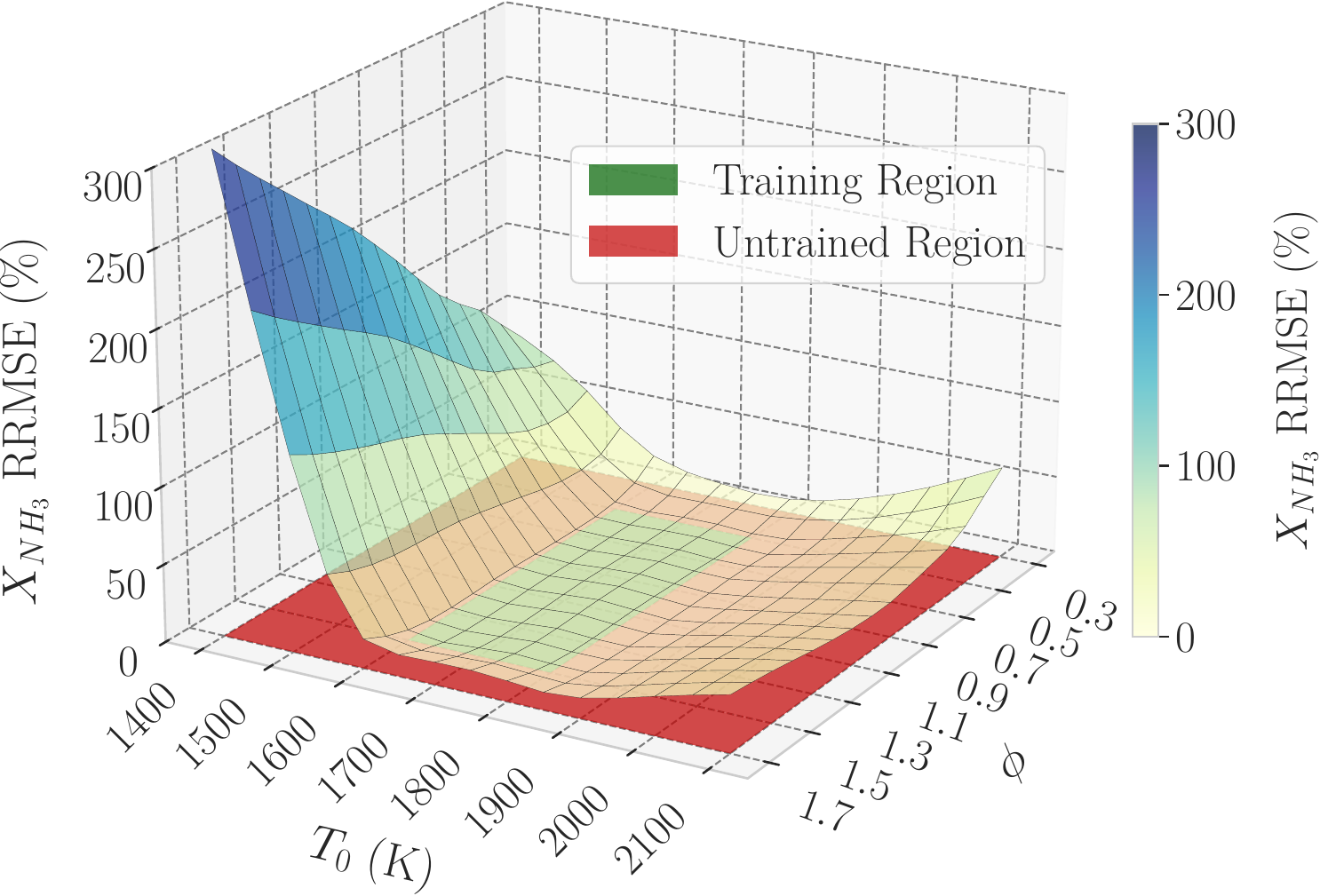}
        \caption{AE+NODE trained with LV.}
        \label{fig:48490_NH3ErrorSurface}
    \end{subfigure}
    \hspace{0.3cm} 
    \begin{subfigure}[b]{0.48\textwidth}
        \centering
        \includegraphics[width=\textwidth]{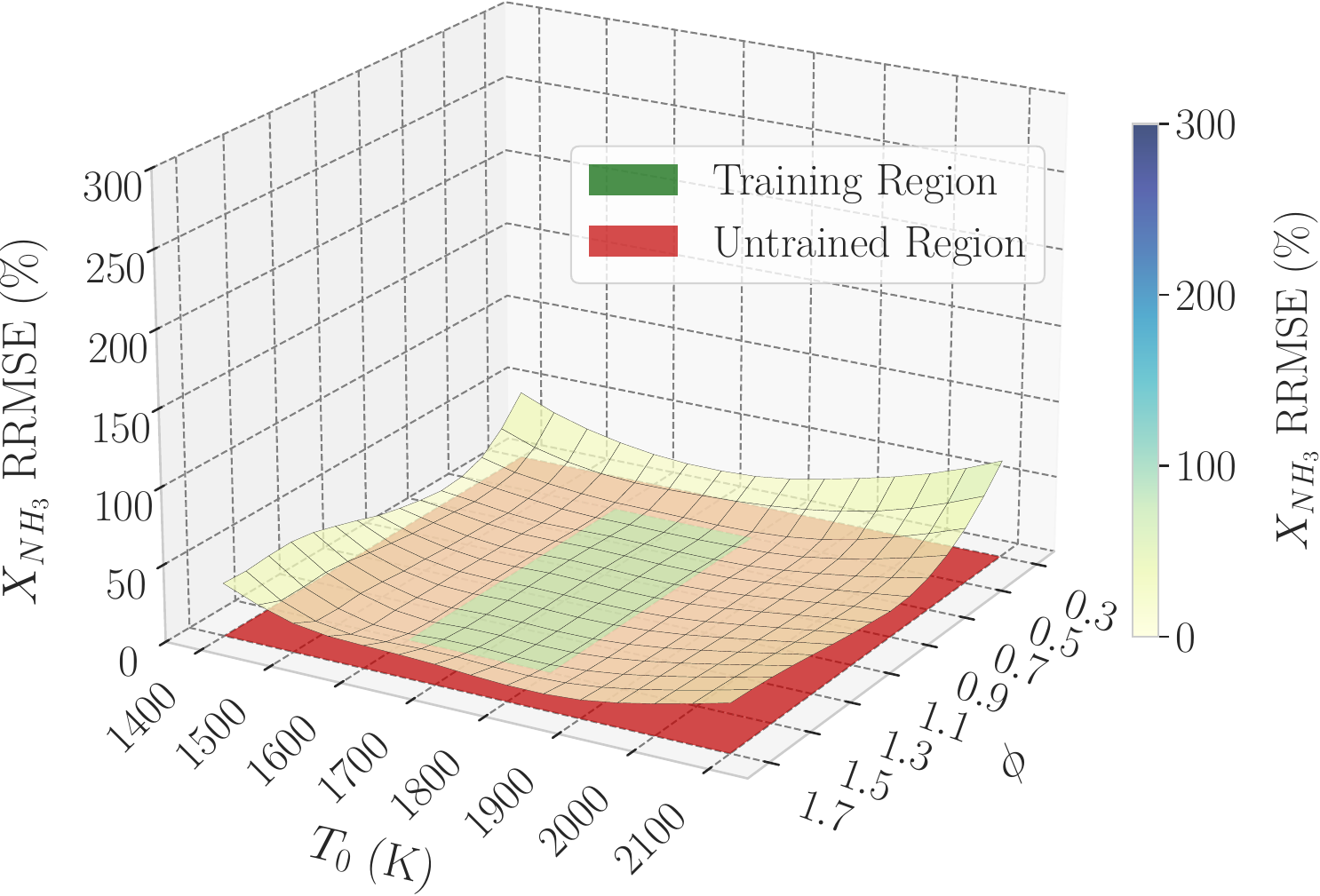}
        \caption{AE+NODE trained with LG.}
        \label{fig:37921_NH3ErrorSurface}
    \end{subfigure}
        \caption{RRMSE for $T$ and $X_{NH_3}$ predictions for conditions within and beyond training range.}
    \label{fig:NH3Mech_Species_RRMSE}
\end{figure}

\begin{figure}[hbt!]

    \centering
    \begin{subfigure}[b]{0.45\textwidth}
        \centering
        \includegraphics[width=\textwidth]{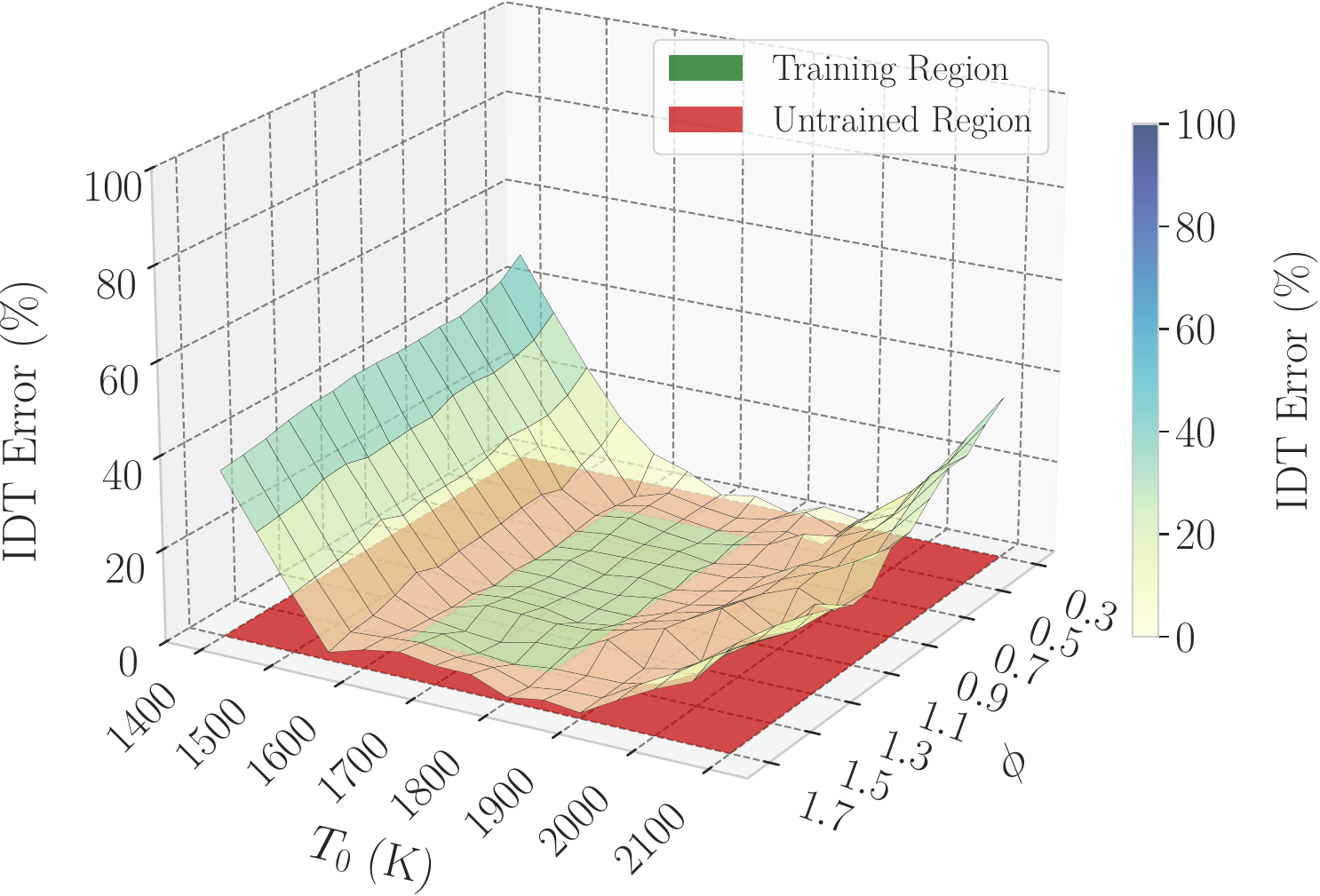}
        \caption{AE+NODE trained with LV.}
        \label{fig:48490_IDTError}
    \end{subfigure}
    \hfill 
    \begin{subfigure}[b]{0.45\textwidth}
        \centering
        \includegraphics[width=\textwidth]{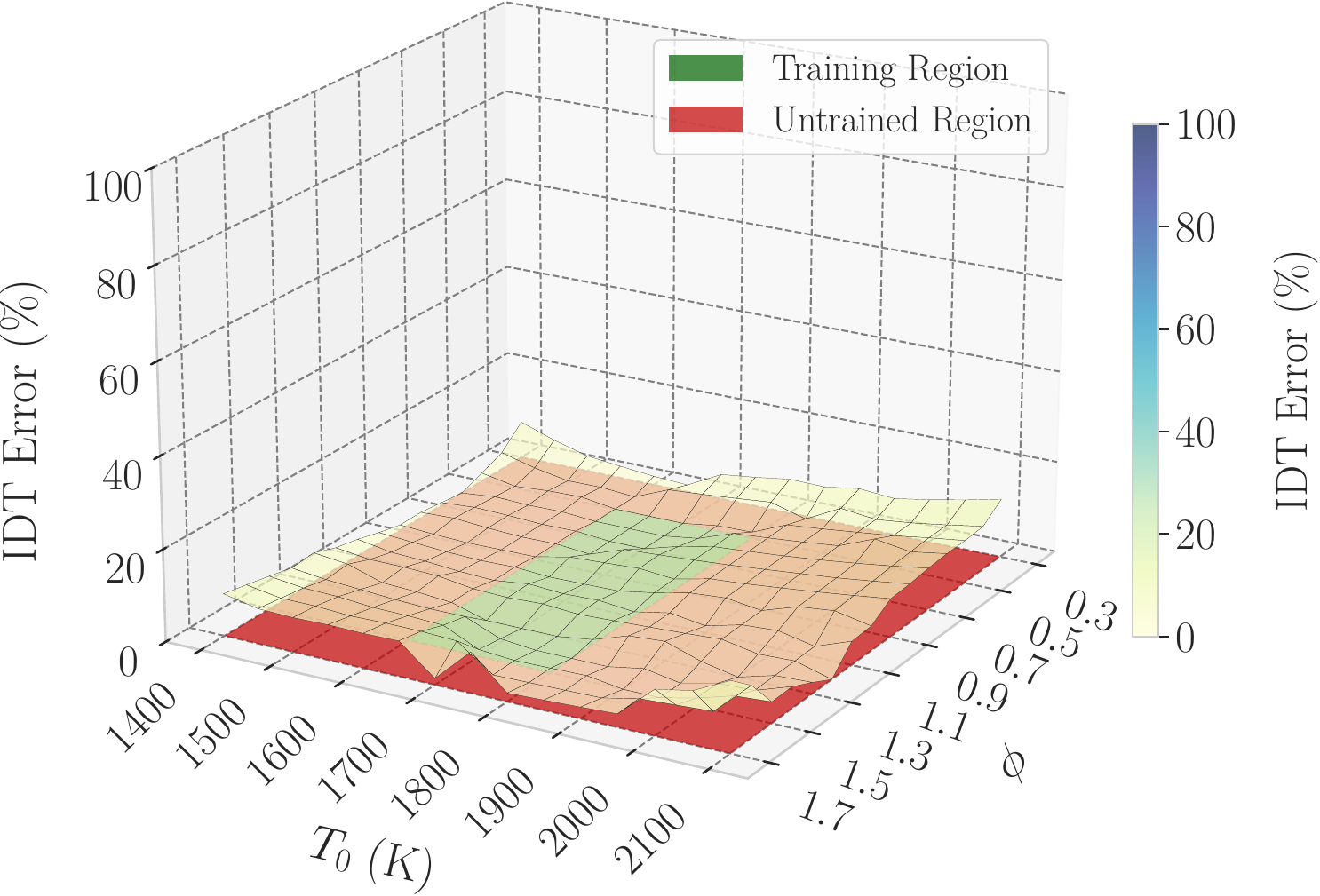}
        \caption{AE+NODE trained with LG.}
        \label{fig:37921_IDTError}
    \end{subfigure}
        \caption{IDT Error for conditions within and beyond the training range.}
    \label{fig:NH3Mech_IDT_Error}
\end{figure}

Fig. \ref{fig:NH3Mech_Species_RRMSE} presents the RRMSE of temperature $T$ and $X_{NH_3}$ for the LV and LG methods for the ammonia/hydrogen-air case. The overall RRMSE appears to be better, not showing the larger errors in the low $T_0$ conditions as seen in the hydrogen-air case (see Fig. \ref{fig:80119_TErrorSurface} and \ref{fig:56908_TErrorSurface}).
Similarly, Fig. \ref{fig:NH3Mech_IDT_Error} shows that the IDT prediction errors for both LV and LG methods are significantly lower compared to those in the hydrogen-air cases. This is because the average IDT for the ammonia/hydrogen-air mixture is an order of magnitude greater than the average IDT for the hydrogen-air mixture. Since the magnitude of the IDT is greater, a similar absolute error in the IDT prediction leads to a smaller relative error. Compared to the LV-trained model, the LG-trained model has a much lower IDT error, in agreement with the results for the hydrogen-air mixture. 

Throughout the extensive comparison of ignition prediction for the hydrogen-air and ammonia/hydrogen-air mixtures, it is concluded that incorporating the latent gradient loss ($L_4$) in the training results in a significant improvement in the AE+NODE prediction capability, specifically for the conditions outside the trained data range. This is attributed to the consideration of a more stringent constraints in the temporal dynamics of the latent variables in the training process.

\FloatBarrier
\subsection{Computational Speed-up}

The AE-NODE is a reduced order model aiming to speed up the computational cost of the prediction while preserving the fidelity. As discussed in \cite{vijayarangan2024data}, the computational speed-up comes in twofold: (a) the reduction of the number of solution variables in the latent space and (b) the removal of stiff time scales in the latent space. The subsection aims to assess how the LV versus LG training methods affect the computational speed-up of the AE+NODE models. Since the dimensionality reduction is predetermined in setting up the autoencoders (by the fixed number of layers and neurons), the analysis first addresses the temporal stiffness reduction, followed by the resultant wall-clock time savings.

The stiffness reduction is quantified by comparing the timesteps required by the explicit time integration in the NODE against those required by the ODE integrator based on Cantera's CVODE\cite{hindmarsh2005sundials}. 
As the AE+NODE architecture and the Cantera solver use different number and forms of the solution variables for integration (latent versus  physical variables), the two simulation cases were run with the same relative tolerance $\varepsilon_\text{rel} = 10^{-4}$ instead of an absolute tolerance.
For each of the hydrogen-air and ammonia/hydrogen-air mixtures, results for representative conditions within and outside the trained conditions are presented. 

\begin{figure}[hbt!]
\centering

    \begin{minipage}{1.0\textwidth} 
    
        \hspace{0.0005\textwidth}
        \begin{subfigure}[b]{0.44\textwidth}
        \raggedright
        \includegraphics[width=\textwidth]{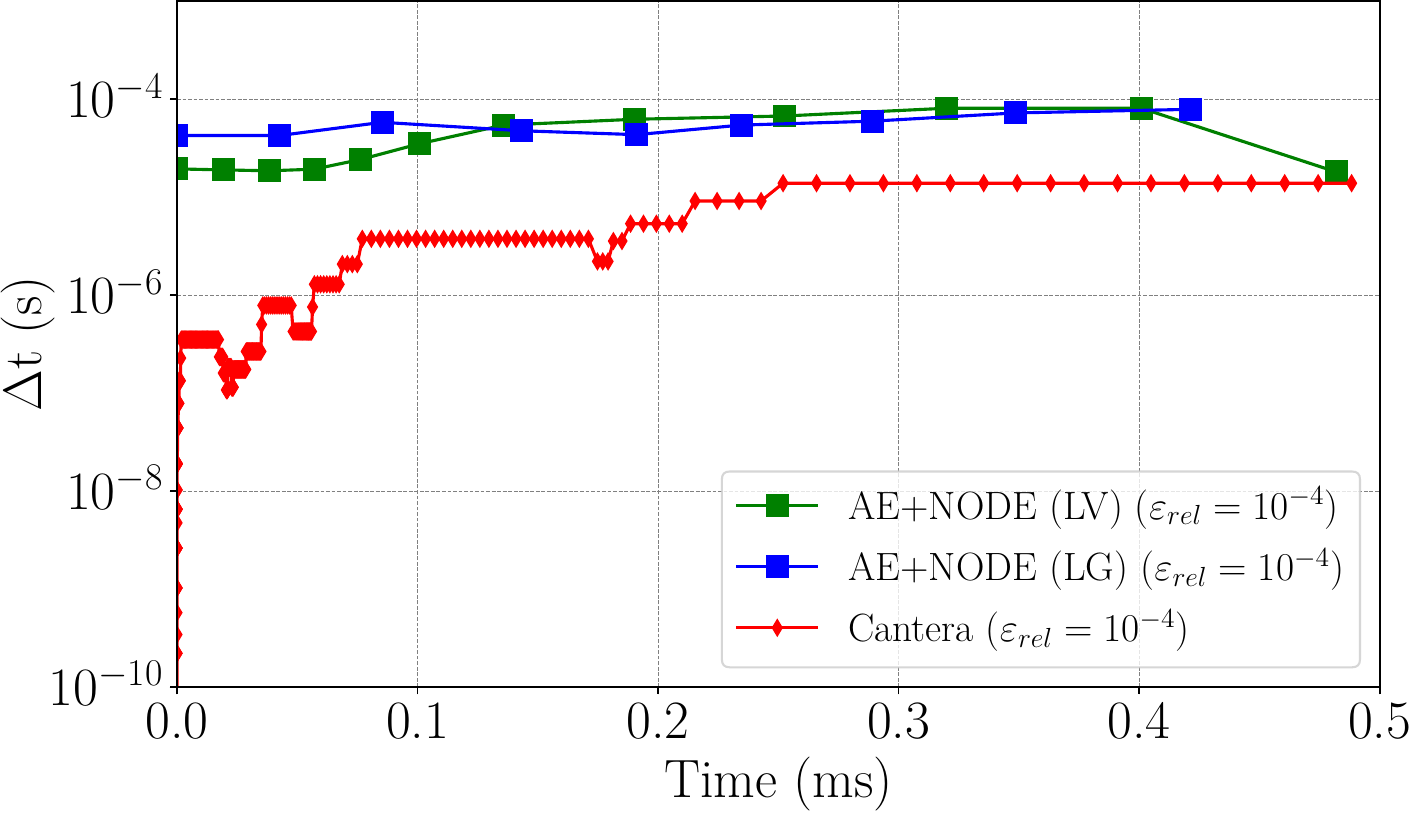}
        \caption{ $\Delta t$ for hydrogen-air mixture at $T_{0}$ = 1300 K, $\phi$ = 1.0.}
        \label{fig:H_AdaptiveT_1300K}
        \end{subfigure}
        \hspace{0.059\textwidth}
        \begin{subfigure}[b]{0.44\textwidth}
        \raggedright
        \includegraphics[width=\textwidth]{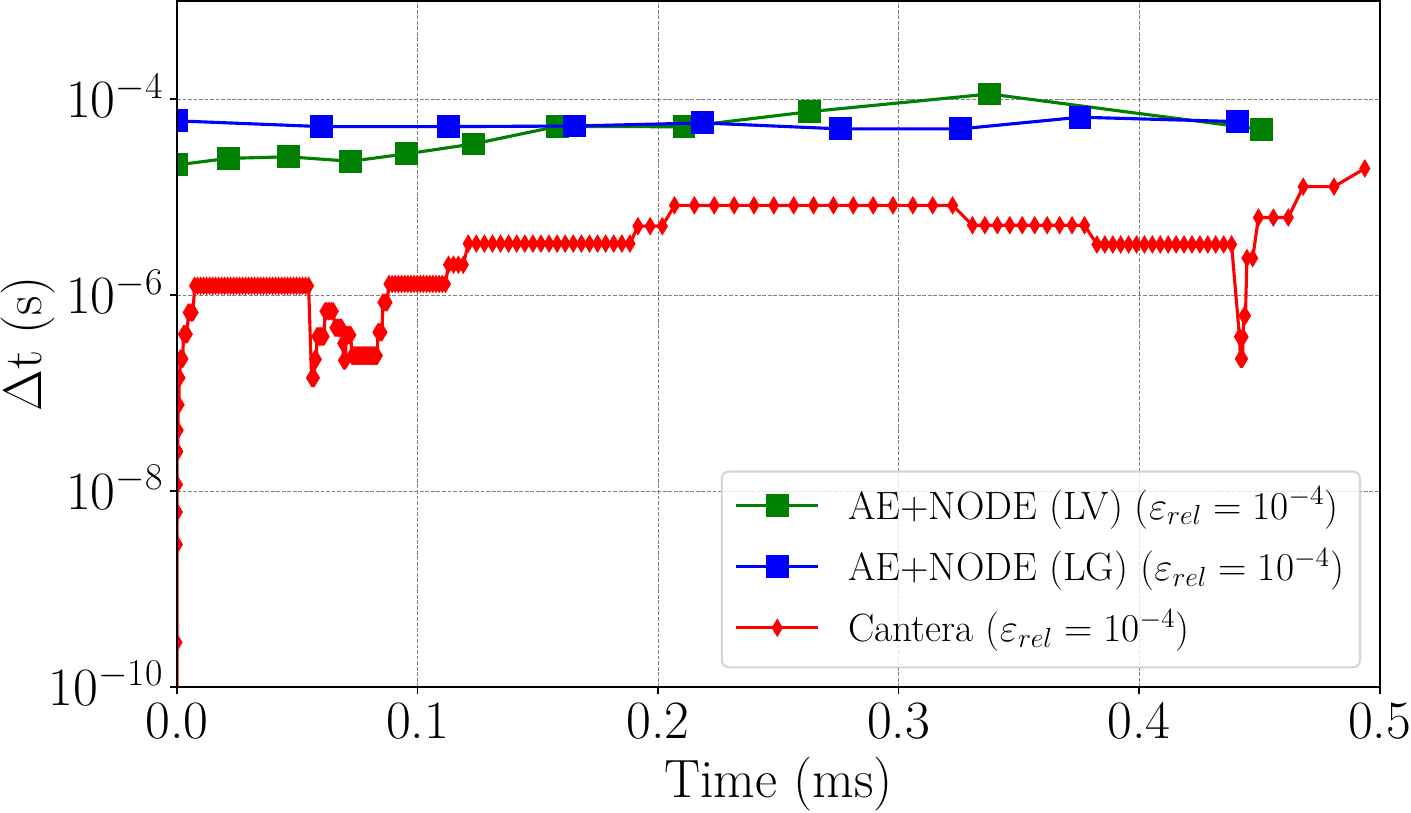}
        \caption{ $\Delta t$ for hydrogen-air mixture at $T_{0}$ = 1100 K, $\phi$ = 1.0.}
        \label{fig:H_AdaptiveT_1100K}
        \end{subfigure}
    \end{minipage}
    \begin{minipage}{1.0\textwidth} 
    \hspace{0.0005\textwidth}
        \begin{subfigure}[b]{0.44\textwidth}
        \raggedright
        \includegraphics[width=\textwidth]{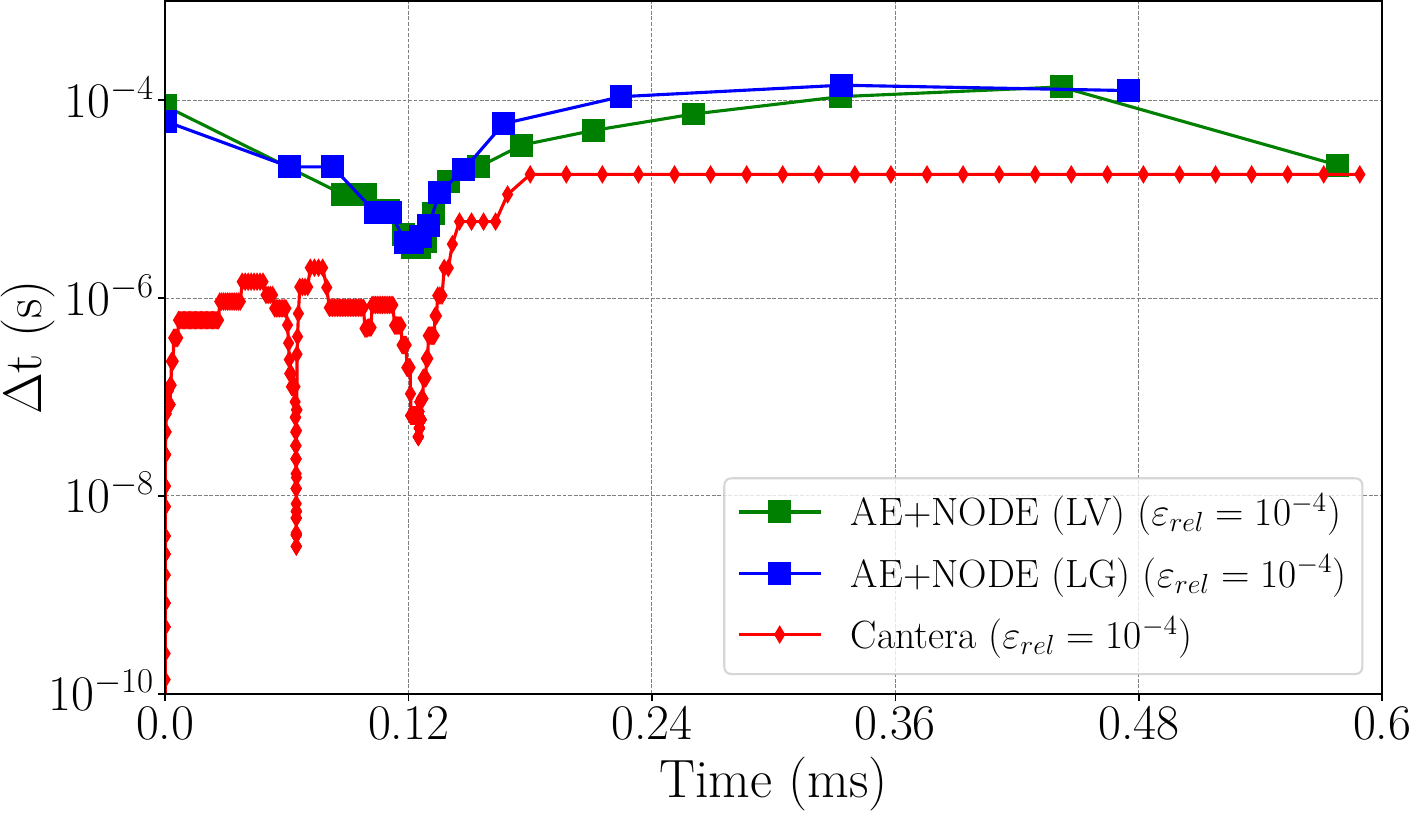}
        \caption{ $\Delta t$ for ammonia/hydrogen-air mixture at $T_{0}$ = 1700 K, $\phi$ = 1.0.}
        \label{fig:NH3_AdaptiveT_1700K}
        \end{subfigure}
        \hspace{0.059\textwidth}
        \begin{subfigure}[b]{0.44\textwidth}
        \raggedright
        \includegraphics[width=\textwidth]{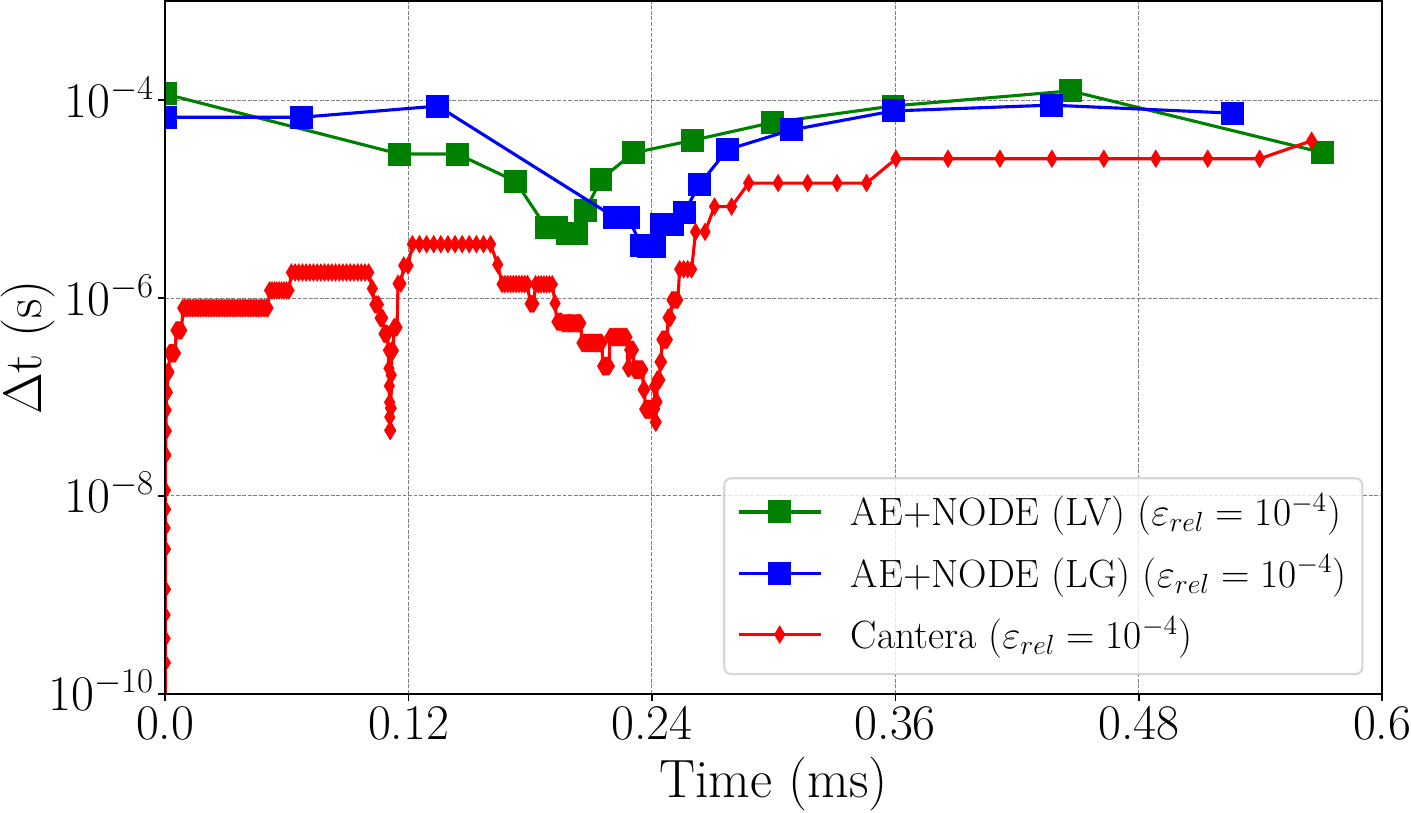}
        \caption{$\Delta t$ for ammonia/hydrogen-air mixture at $T_{0}$ = 1500 K, $\phi$ = 1.0.}
        \label{fig:NH3_AdaptiveT_1500K}
        \end{subfigure}
    \end{minipage}
    \caption{Comparison of computational timesteps throughout the simulations for representative cases.}
    \label{fig:AdaptiveT_all}
\end{figure}

Fig. \ref{fig:AdaptiveT_all} shows the comparison of time steps throughout the simulations for the hydrogen-air (a) within trained data ($T_0 = 1300$K), (b) outside the trained data  ($T_0 = 1100$K), and for the ammonia/hydrogen-air (c) within trained data ($T_0 = 1700$K), and (d) outside the trained data  ($T_0 = 1500$K). For all cases, AE-NODE achieves a consistent increase in the timestep by one to two orders of magnitude throughout the simulation. For the hydrogen-air case, the temporal stiffness is uniformly removed even during the ignition event. For the ammonia/hydrogen-air cases, some reductions in the timestep occurs with AE-NODE, possibly due to some fast chemical timescales associated with the ammonia chemistry. Nevertheless, the abrupt changes in timesteps as seen in the Cantera integration is absent in the AE-NODE model. While the LG method was found to improve the solution accuracy, the stiffness removal performance does not necessarily lead to a noticeable improvement over the LV method, and the timesteps are found to be comparable for all cases under study.

Finally, the net computational speed-up is assessed by the wall-clock computational time for the AE+NODE and Cantera solutions, both using the same programming language, C++. The AE+NODE architecture was implemented in C++ by using Odeint \cite{ahnert_odeint-solving_2011}, and Eigen \cite{eigenweb} libraries. Similarly, the open-source C++ version of Cantera was used. 

The wall-clock time comparison was made in two different ways: (i) adaptive time step integration with a fixed relative tolerance ($\varepsilon_\text{rel}$) of $10^{-4}$, and (ii) constant time step integration using the minimum time step from the adaptive time step integration. The motivation for the second method is to represent the scenarios in practical 2D or 3D simulations, where a constant timestep is often chosen by the global minimum values across the flame in order to ensure stability. 
Hence, the second test may be relevant when the AE+NODE architecture is implemented in 2D or 3D simulations.  For all cases under study, simulations were repeated 100 times to obtain reliable averages. Only the integration part of the simulations was timed. 

\begin{figure}[hbt!]
\centering
    \begin{subfigure}[b]{0.49\textwidth}
    \centering
        \includegraphics[width=\textwidth]{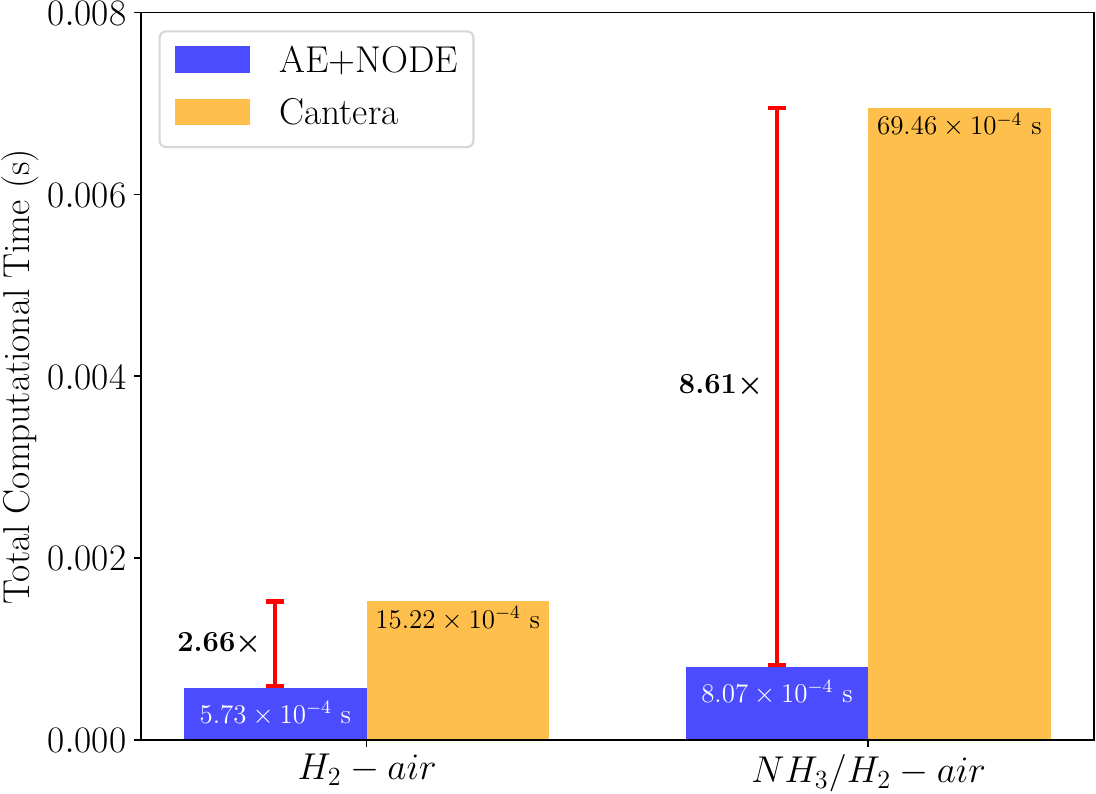}
        \caption{Adaptive Time Step}
        \label{fig:Speedup_method1}
    \end{subfigure}
    \hfill
    \begin{subfigure}[b]{0.49\textwidth}
    \centering
        \includegraphics[width=\textwidth]{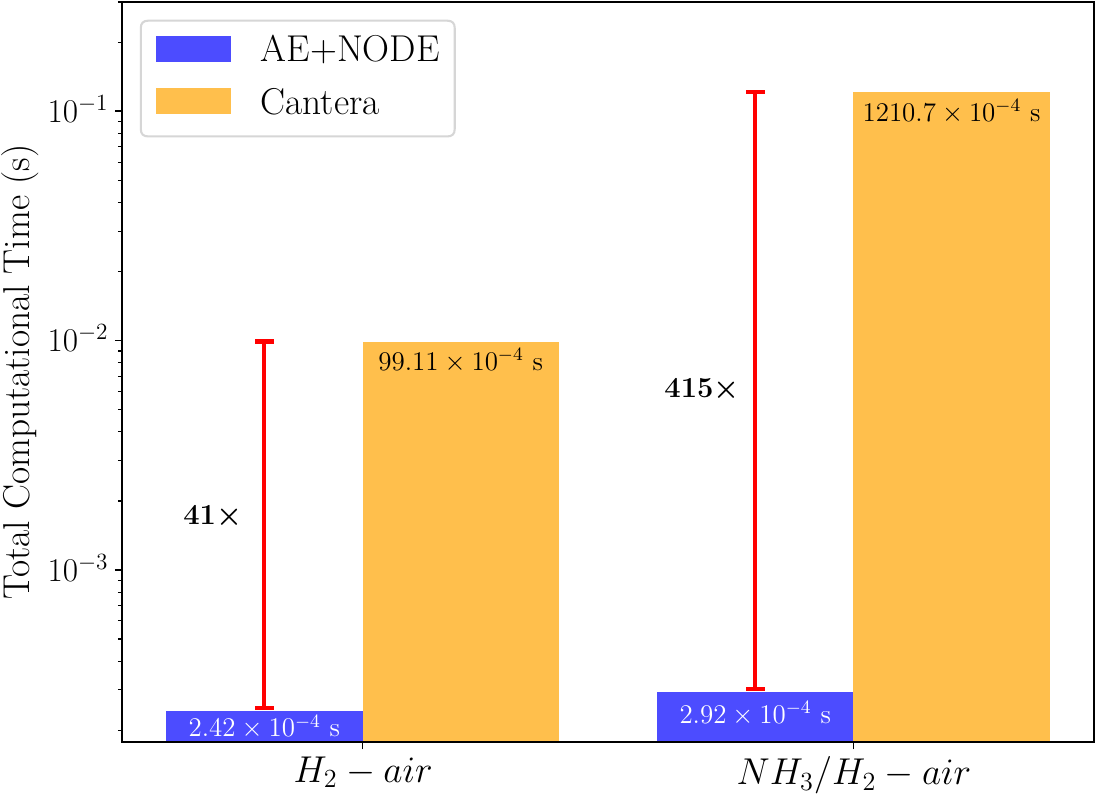}
        \caption{Fixed Time Step}
        \label{fig:Speedup_method2}
    \end{subfigure}
    \caption{AE+NODE vs Cantera, the total computational time for integration.}
\end{figure}

Fig. \ref{fig:Speedup_method1} shows the wall-clock times with the first method. The AE+NODE speed-up was $2.66$X for the hydrogen-air and   $8.61$X for the ammonia/hydrogen-air mixture, respectively. Note that, for the AE+NODE, the total computational time for integrating the ammonia/hydrogen-air mixture is only $1.41$X that of integrating the hydrogen-air mixture. On the other hand, the total computational time for Cantera integration of the ammonia/hydrogen-air mixture is $4.57$X that of the hydrogen-air mixture.
This may be attributed to the fact that the same autoencoder dimension was used for both fuel mixture cases. The selection of the number of neurons and layers is mostly done by empirical trial and error, and more work is needed to find an optimal dimension of the autoencoder in the reduced order model development.
It is also noted that the speed-up achieved with the AE+NODE architecture was found to be comparable for the LV and the LG method, as expected from the timestep analysis shown in Fig. \ref{fig:AdaptiveT_all}.

Fig. \ref{fig:Speedup_method2} shows the speed-up for the second scenario of fixed timesteps. 
The speed-up by the AE-NODE is significantly larger in this case, with $41$X for the hydrogen-air and  $415$X for the ammonia/hydrogen-air cases, respectively.
On average, the minimum $\Delta t$ for the AE+NODE is on the scale of $10^{-5}$ for both hydrogen and ammonia/hydrogen-air mixtures, but for Cantera, the minimum $\Delta t$ is on the scale of $10^{-8}$ and $10^{-9}$ respectively.  For the hydrogen-air mixture, AE+NODE speedup was $41$X, and for the ammonia/hydrogen-air mixture, it was $415$X. This results reflects the findings in Fig. \ref{fig:AdaptiveT_all}, implying that the dynamic range of the timestep variation throughout the entire ignition event is much more compressed with the AE-NODE, unlike the Cantera case where rapid and drastic changes in timesteps occur during the event. This further demonstrates a potential advantage of the AE+NODE for stiff chemical systems in multi-dimensional simulations.

\label{fig:Speedup_method_combined}

\FloatBarrier
\section{Conclusion} \label{sec:Conclusion}

This study introduces a latent gradient loss term ($L_4$) into the training strategy of an autoencoder (AE) coupled with a Neural Ordinary Differential Equation (NODE). Compared to the commonly used latent variable (dynamics-informed) training approach \cite{vijayarangan2024data}, the proposed method leads to more robust models for both datasets under study. The proposed methodology demonstrates improved generalization and increased robustness for predicting the evolution of the thermochemical state variables and ignition delay times, particularly for conditions beyond the training range. However, these improvements come at the cost of significantly longer training times and lower training accuracy at saturation.

Through integration in the latent space, the AE+NODE architecture  achieves a significant computational advantage compared to conventional stiff solvers ssuch as the backward differentiation formula (BDF) employed in the CVODE algorithm in Cantera. The AE+NODE further allows a significant reduction in temporal stiffness, leading increase in the integration timestep by approximately $O(10^3)$, while maintaining high accuracy and demonstrating better scalability. The level of computational speedup is found to be consistent, independent of the specific training loss used.

The results imply that for limited or sparse datasets, utilizing the more stringent latent gradient loss term during training leads to improved model generalization capability. Moreover, the demonstrated reduction in integration time indicates that the AE+NODE model has the potential to significantly accelerate chemistry computations required in large scale reactive flow simulations.

Further research is required to expand the model to cover a broader range of thermochemical states, enabling the AE+NODE architecture to approximate the full range of reactions contained in a chemical kinetics mechanism to an accuracy acceptable for CFD simulations. Additionally, the integration of the AE+NODE framework for source term evaluation in multidimensional reactive flow simulations, such as 2D and 3D flame configurations in platforms like OpenFOAM \cite{jasak2007openfoam}, needs to be demonstrated.

\section*{Acknowledgments}

This work was supported by the Korea Advanced Institute of Science and Technology (KAIST) Presidential Fellowship Program (KPF) and the King Abdullah University of Science and Technology (KAUST), in particular the Visiting Student Research Program (VSRP) for the first author.

\section*{Appendix}
\appendix
\setcounter{figure}{0}
\renewcommand{\thefigure}{A.\arabic{figure}}
\subsection*{Appendix A: Ammonia/hydrogen-air mixture, accuracy in trained conditions}\label{sec:AppendixA}

\begin{figure}[hbt!]
    \centering
    \begin{subfigure}[b]{0.42\textwidth}
        \centering
        \includegraphics[width=\textwidth]{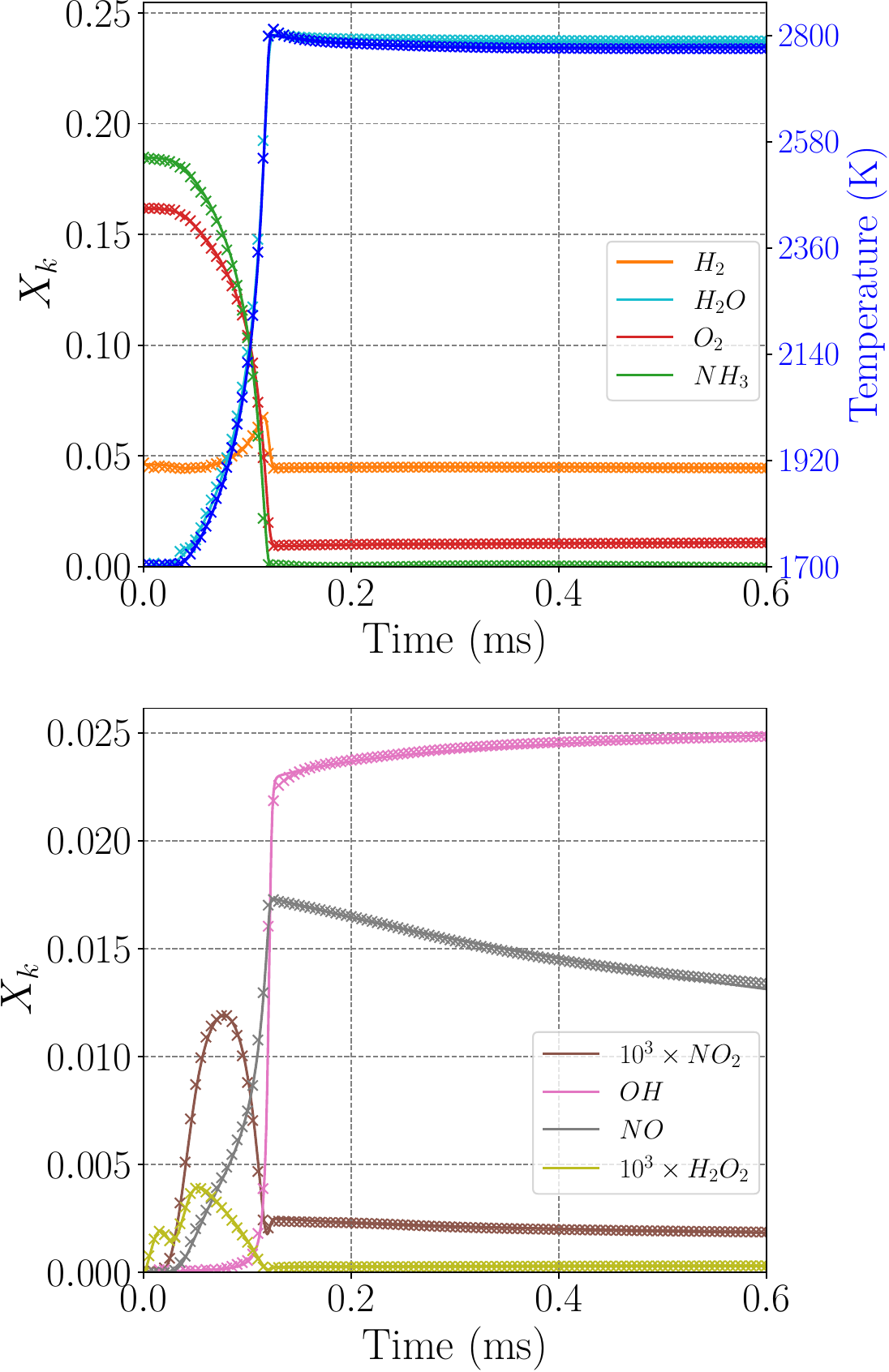}
        \caption{AE+NODE trained with LV.}
        \label{fig:48490_T1700P1P1}
    \end{subfigure}
    \hspace{0.3cm}
    \begin{subfigure}[b]{0.42\textwidth}
        \centering
        \includegraphics[width=\textwidth]{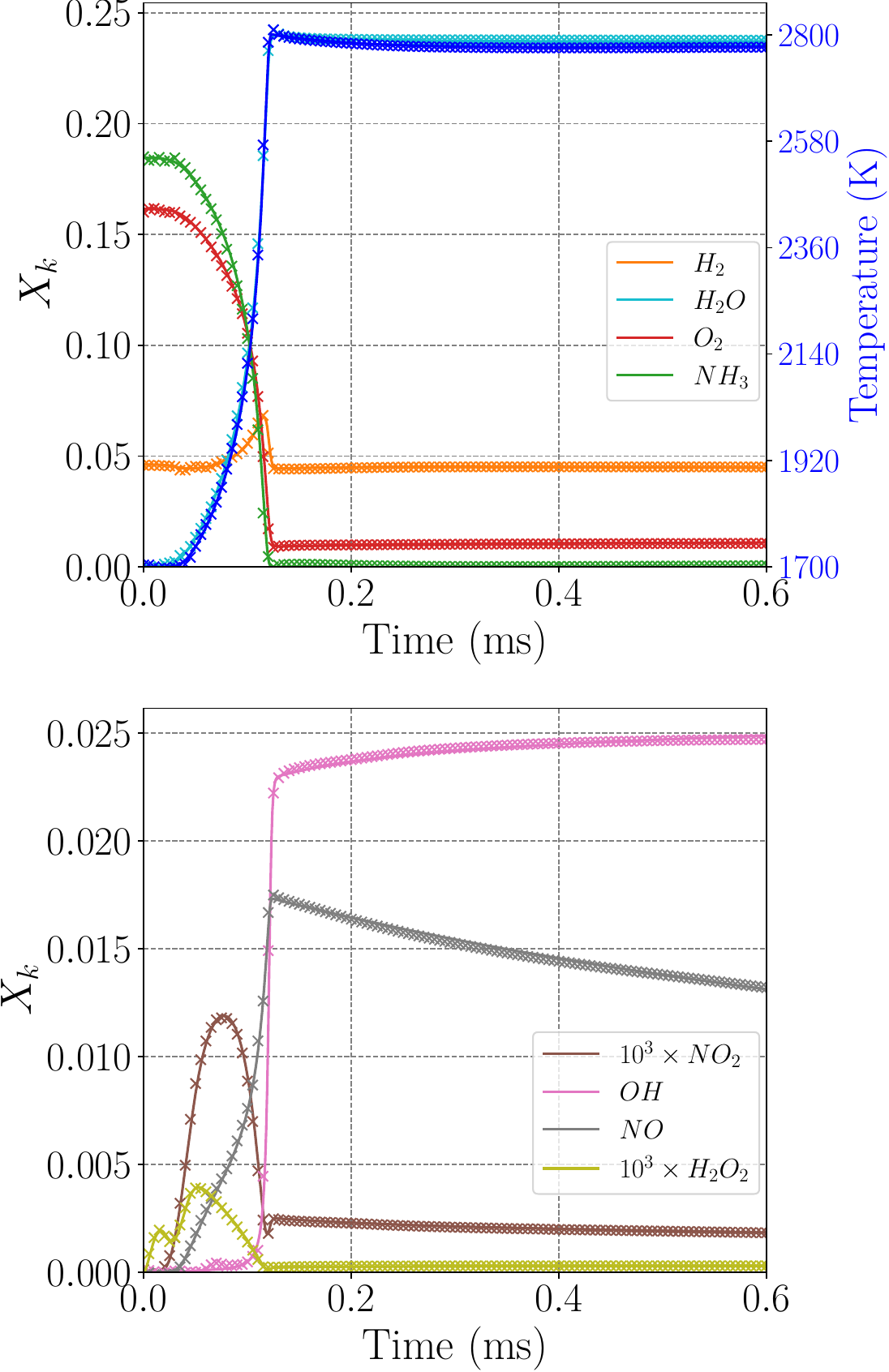}
        \caption{AE+NODE trained with LG.}
        \label{fig:37921_T1700P1P1}
    \end{subfigure}
    \caption{AE+NODE (symbols) and Cantera (lines) predictions of the physical state variables during constant-pressure homogeneous ignition for a trained condition \(T_{0}\) = 1700 K, \(\phi = 1.0\).}
    \label{fig:NH3_Validation_1700K}
\end{figure}

Fig. \ref{fig:NH3_Validation_1700K} shows the comparison of AE-NODE predictions against the reference Cantera solutions of temporal evolution of temperature and major/minor species mole fractions. With the level of training the model comparable to that in the hydrogen-air case, excellent predictions are observed for all solution variables for both LV and LG methods for the initial conditions within the training range. No additional complexities associated with the larger dimensionality of the ammonia/hydrogen chemistry are found.

\begin{figure}[hbt!]
        \centering
        \includegraphics[height=0.25\textheight]{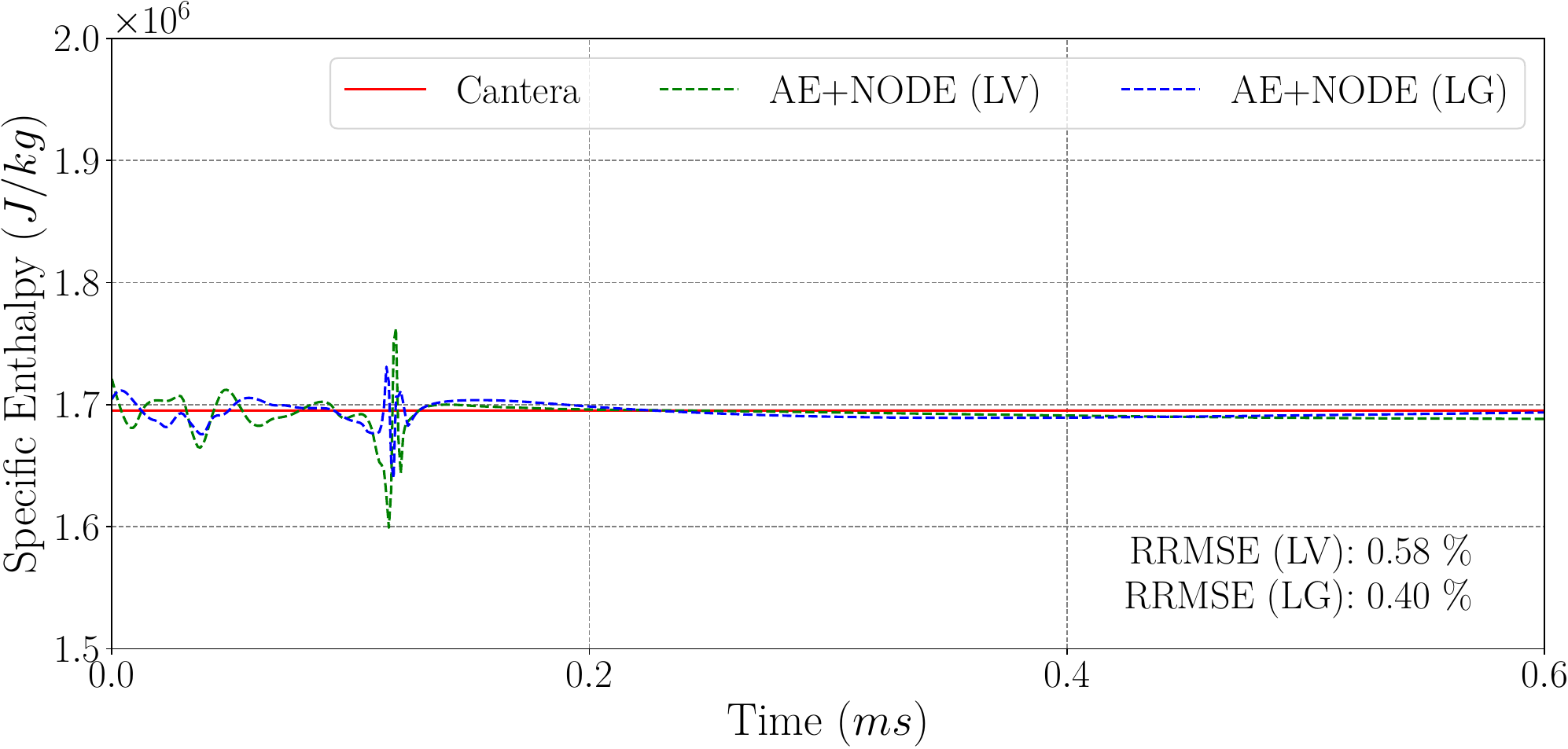}
        \caption{Evolution of the total specific enthalpy during constant-pressure homogeneous ignition (T$_{0}$ = 1700 K, $\phi$ = 1.0, and $X_{NH_{3}}$:$X_{H_{2}}$$ = 0.8:0.2$).}
        \label{fig:NH3Mech_T1700K_Enthalpy}
\end{figure}

Fig. \ref{fig:NH3Mech_T1700K_Enthalpy} further shows the evolution of the total specific enthalpy for both LV and LG methods, confirming that the enthalpy conservation is satisfied with the maximum errors during ignition at the comparable level as those found in the hydrogen-air system, and RRMSE values maintained below $1\%$.

\begin{figure}[hbt!]
    \centering
    \begin{subfigure}[b]{0.42\textwidth}
        \centering
        \includegraphics[width=\textwidth]{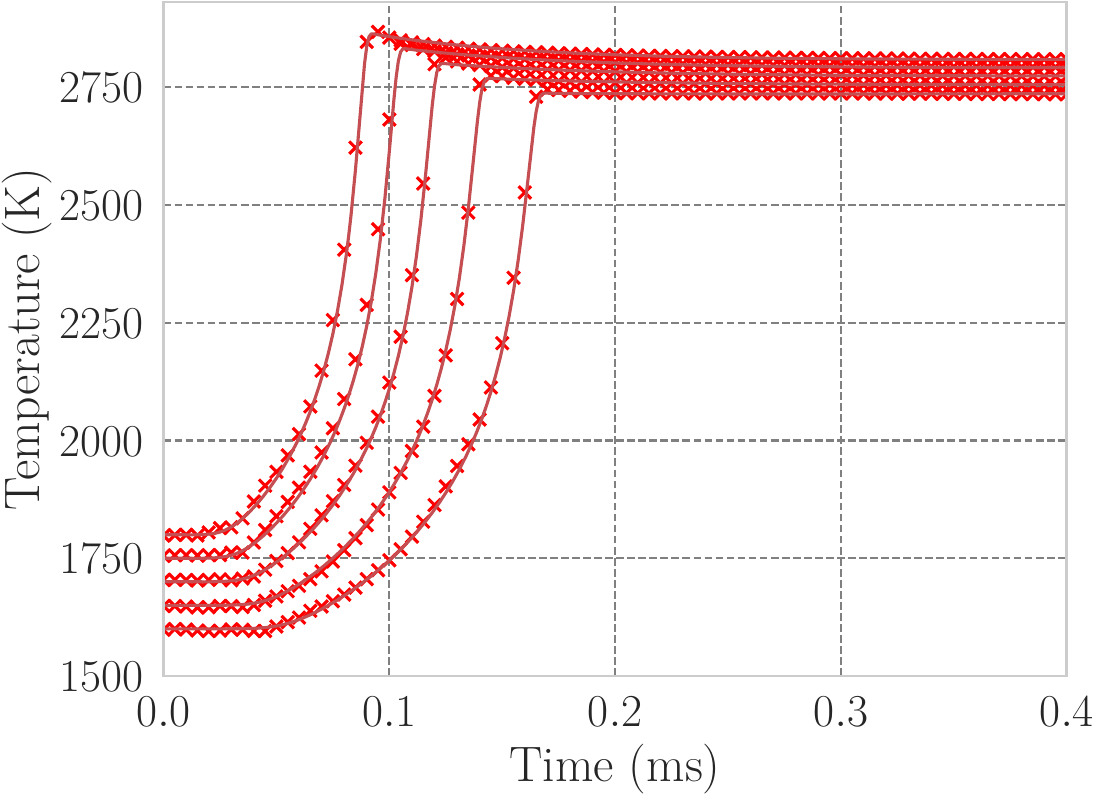}
        \caption{AE+NODE trained with LV.}
        \label{fig:48490_TempTraj}
    \end{subfigure}
    \hspace{0.3cm} 
    \begin{subfigure}[b]{0.42\textwidth}
        \centering
        \includegraphics[width=\textwidth]{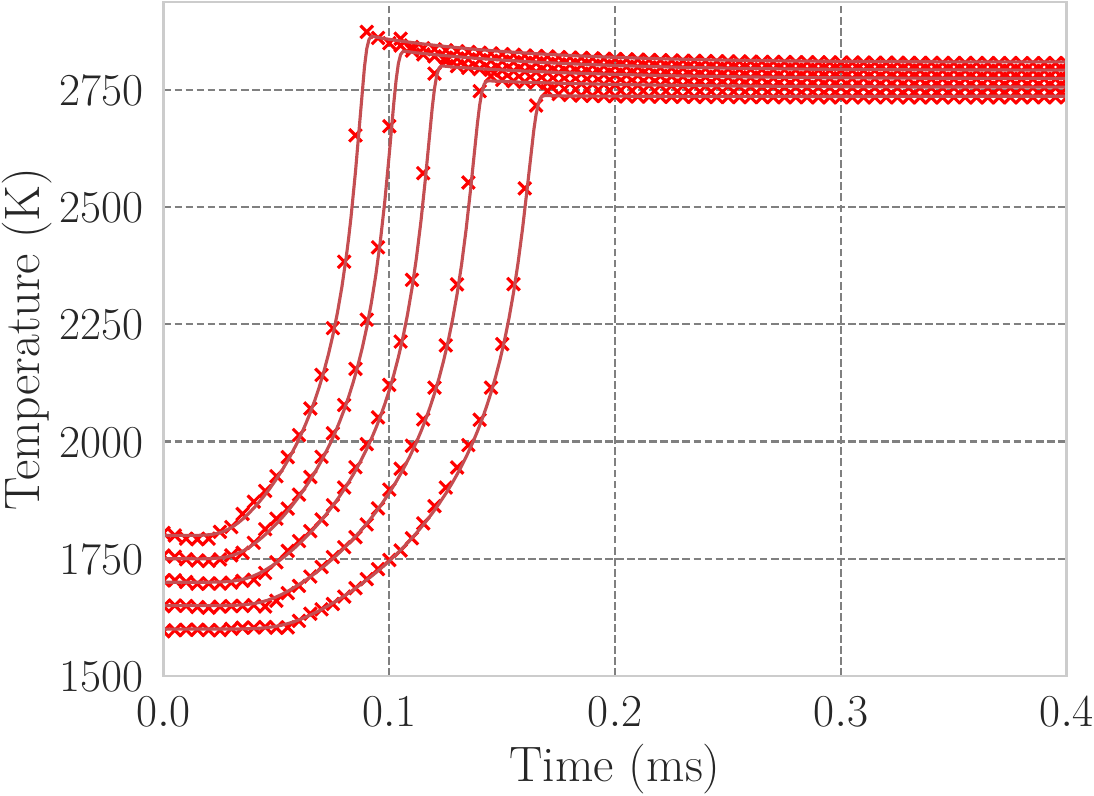}
        \caption{AE+NODE trained with LG.}
        \label{fig:37921_TempTraj}
    \end{subfigure}
        \caption{AE+NODE ($\mathsf{x}$) vs Cantera (line), the evolution of temperature over time for various $T_0$ at $\phi = 1.0$.}
    \label{fig:NH3Mech_Temp_Trajectories_train}
\end{figure}

Predicted temperature evolutions by the LV and LG models for initial conditions within the training data range at a constant equivalence ratio are compared in Fig. \ref{fig:NH3Mech_Temp_Trajectories_train}. As with the hydrogen-air mixture, both models accurately capture the temperature evolution for the trained and interpolated conditions.

\FloatBarrier
\setcounter{figure}{0}
\renewcommand{\thefigure}{B.\arabic{figure}}
\subsection*{Appendix B: Ammonia/hydrogen-air mixture, accuracy outside trained conditions}\label{sec:AppendixB}

\begin{figure}[hbt!]
\centering
   \begin{subfigure}[b]{0.42\textwidth}
        \centering
        \includegraphics[width=\textwidth]{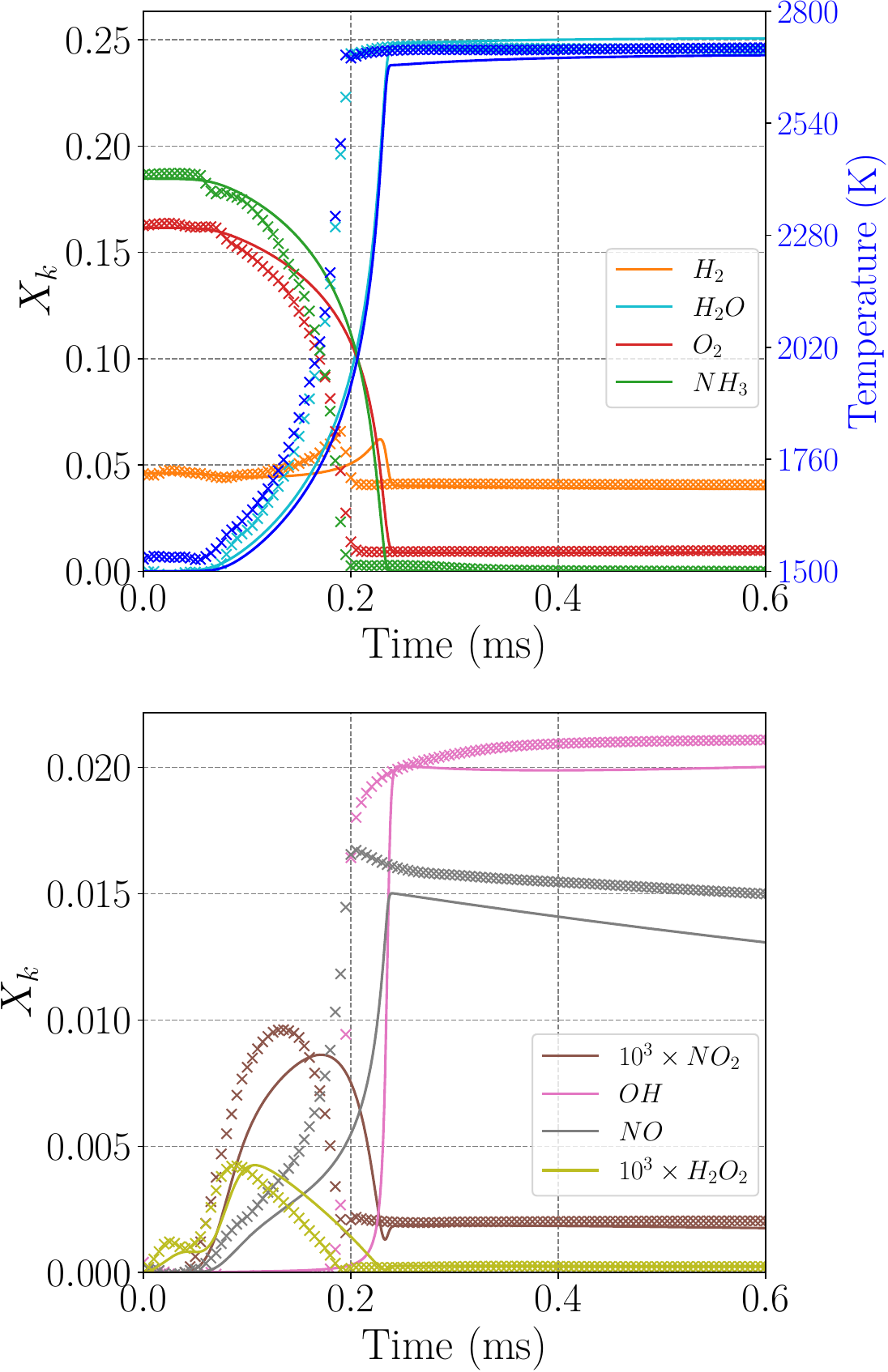}
        \caption{AE+NODE trained with LV}
        \label{fig:NH3Mech_T1500K_Comp_LV}
    \end{subfigure}
    \hspace{0.3cm} 
    \begin{subfigure}[b]{0.42\textwidth}
        \centering
        \includegraphics[width=\textwidth]{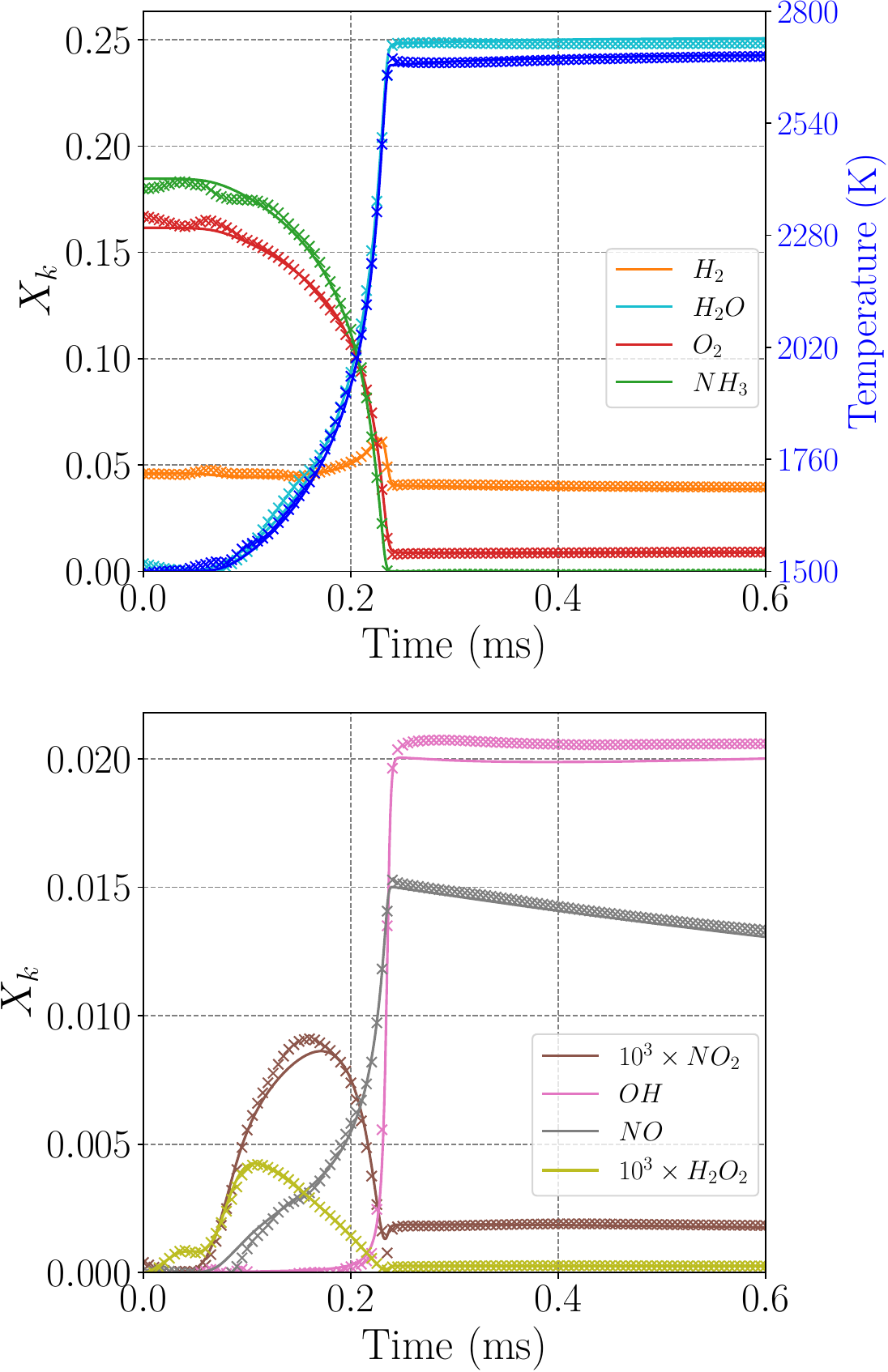}
        \caption{AE+NODE trained with LG}
        \label{fig:NH3Mech_T1500K_Comp_LG}
    \end{subfigure}
        \caption{AE+NODE (symbols) and Cantera (lines) predictions of the time evolution of physical state variables during constant-pressure homogeneous ignition for an untrained condition (T$_{0}$ = 1500 K, $\phi$ = 1.0, and $X_{NH_{3}}$:$X_{H_{2}}$$ = 0.8:0.2$.)}
    \label{fig:NH3Mech_T1500K_Comp}
\end{figure}

Fig. \ref{fig:NH3Mech_T1500K_Comp} shows the evolution of selected species and temperature for the ammonia/hydrogen-air reaction for initial conditions of $T_\text{0} = 1500 \text{ K}$ and $\phi = 1.0$, which are out of the training range. As in the hydrogen-air case, the LV-trained model predicts the ignition event earlier by approximately 0.04 ms, evidenced by the leftward shift of temperature and species predictions in Fig. \ref{fig:NH3Mech_T1500K_Comp_LV}. The underprediction of the ignition delay time is related to the overprediction of the initial temperature being off by 50.8 K as shown in Fig. \ref{fig:NH3Mech_T1500K_Comp_LV}. In contrast, the LG-trained model accurately predicts the initial temperature and ignition delay. Due to the complexity of the system, however, the agreement between the AE+NODE and reference predictions is not as good as that found in the hydrogen-air case. 

\begin{figure}[hbt!]
        \centering
        \includegraphics[height=0.25\textheight]{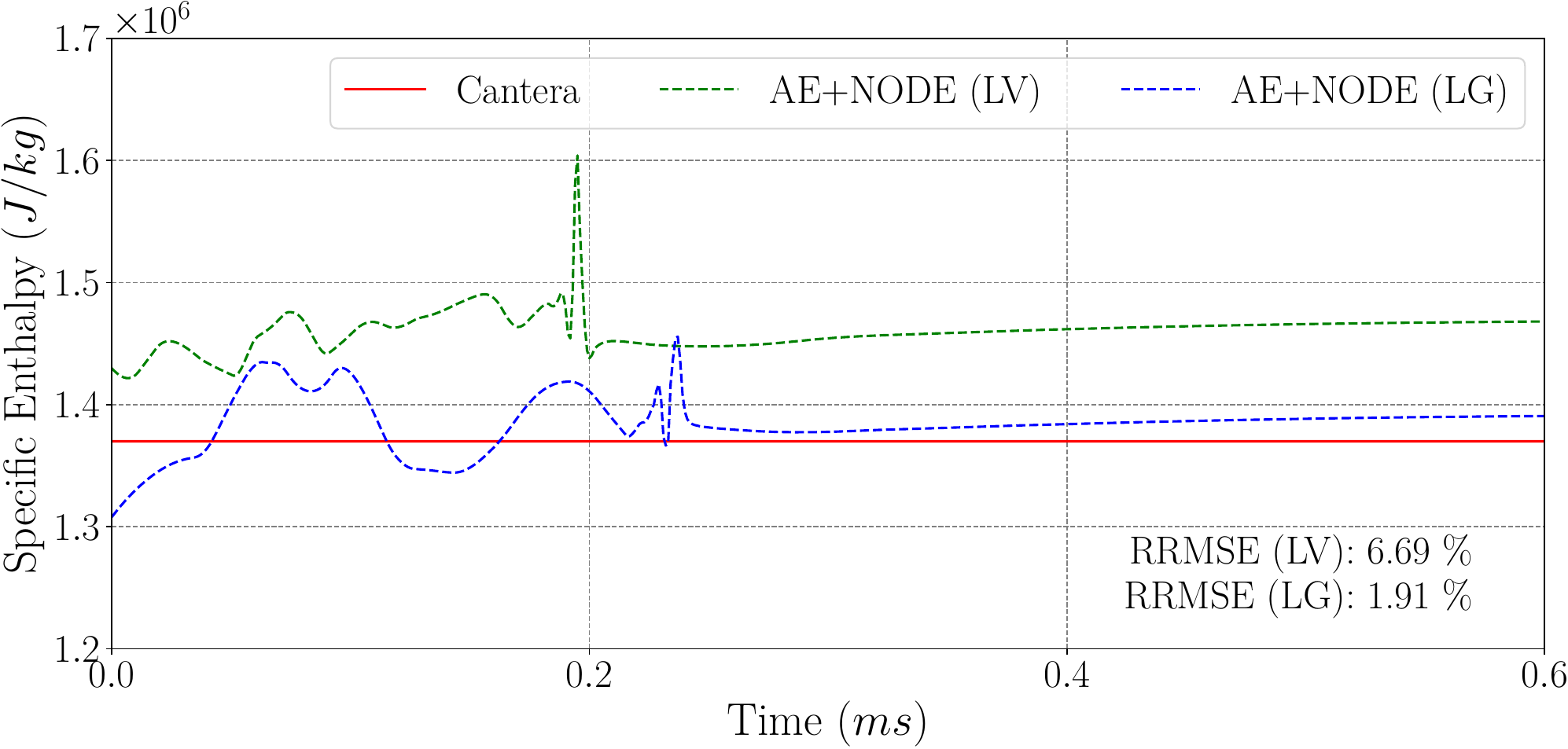}
        \caption{Evolution of the total specific enthalpy during constant-pressure homogeneous ignition (T$_{0}$ = 1500 K, $\phi$ = 1.0, and $X_{NH_{3}}$:$X_{H_{2}}$$ = 0.8:0.2$).}
        \label{fig:NH3Mech_T1500K_Enthalpy}
\end{figure}

The evolution of the total specific enthalpy shown in Fig. \ref{fig:NH3Mech_T1500K_Enthalpy} further reveals the degradation of the prediction fidelity in enthalpy conservation, which should remain constant at $1.37\times10^6 \text{ J}/\text{kg}$. While the level of discrepancy for the LV method is similar to that in the hydrogen-air case, even the LG method reveals noticeable errors in the early phase up to ignition with significant fluctuation, while the terminal value after ignition is also slightly overpredicted. For this condition, the LG method results in the RRMSE of 1.91\%, higher than that in the hydrogen-air case at 1.46\%. Nevertheless, the errors in the predictions of physical state variables, and thus specific enthalpy, are much lower than those of the LV-trained model, in line with the results observed for the hydrogen-air mixture.

\begin{figure}[hbt!]
    \centering
    \begin{subfigure}[b]{0.42\textwidth}
        \centering
        \includegraphics[width=\textwidth]{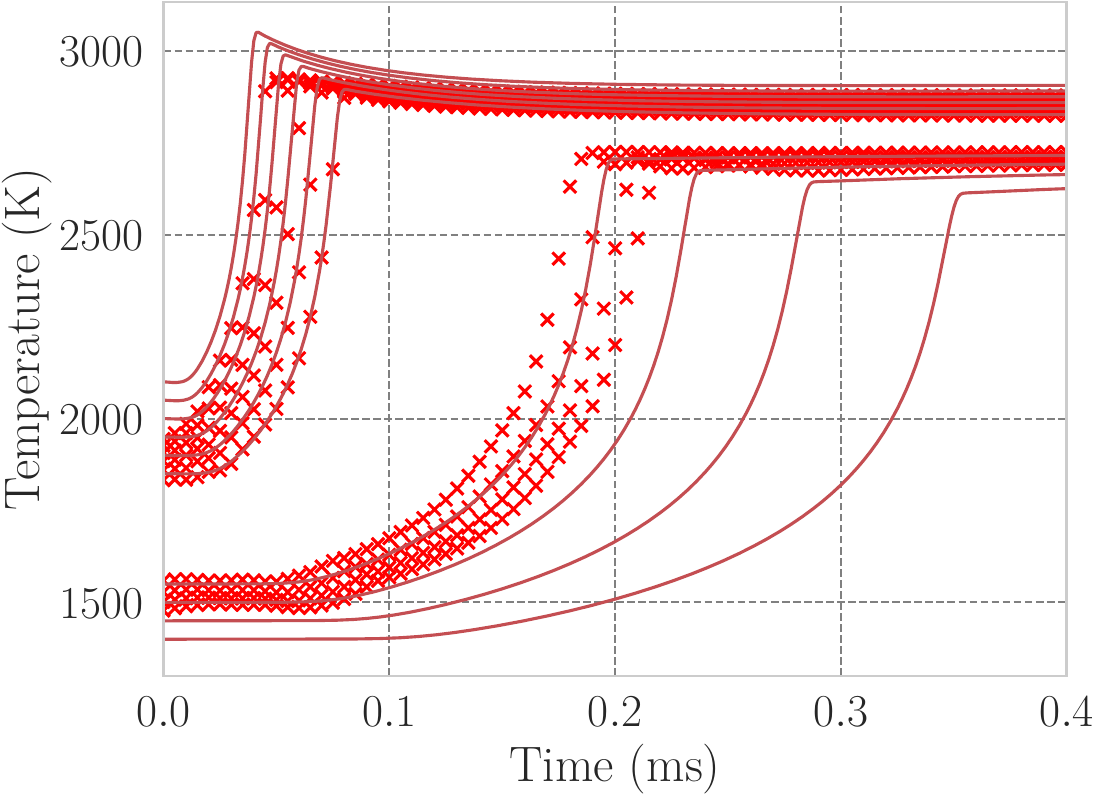}
        \caption{AE+NODE trained with LV.}
        \label{fig:48490_TempTraj_v2}
    \end{subfigure}
    \hspace{0.3cm} 
    \begin{subfigure}[b]{0.42\textwidth}
        \centering
        \includegraphics[width=\textwidth]{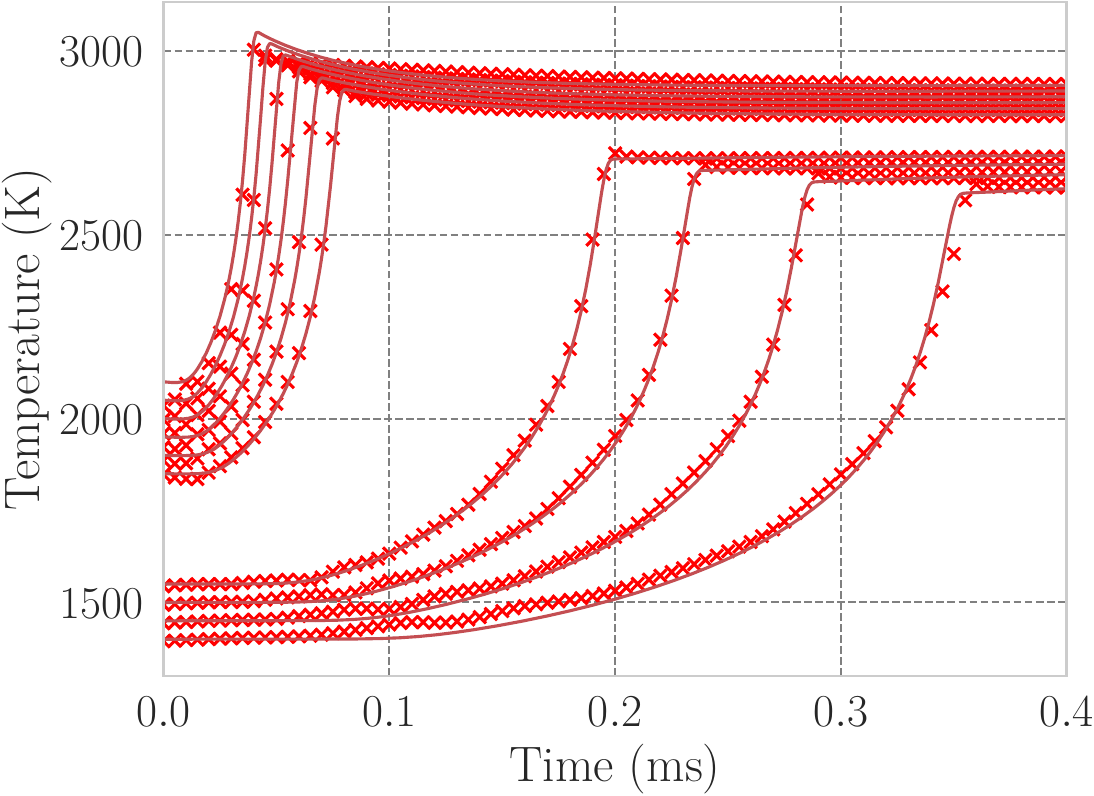}
        \caption{AE+NODE trained with LG.}
        \label{fig:37921_TempTraj_v2}
    \end{subfigure}
        \caption{AE+NODE ($\mathsf{x}$) vs Cantera (line), the evolution of temperature over time for various $T_0$ at $\phi = 1.0$.}
    \label{fig:NH3Mech_Temp_Trajectories_test}
\end{figure}

The predicted temperature evolution, outside the training range shown in Fig. \ref{fig:NH3Mech_Temp_Trajectories_test} is found to be consistent with the results in the hydrogen-air case (Fig. \ref{fig:H2Mech_Temp_Trajectories_Test}), in that the LV method yields significant errors in the initial temperature and the overall dynamics as the condition deviates away from the trained range, while the LG method yields accurate predictions. It is noted that the initial errors with the LV method at higher temperature conditions differ between the hydrogen-air case (overprediction) and the ammonia/hydrogen-air case (underprediction).

\bibliography{elsarticle-template}

\end{document}